\documentclass[twoside,12pt]{article}
\usepackage{CJK}
\usepackage{indentfirst}
\usepackage{bm}
\usepackage{epsfig}
\usepackage{mathptmx}
\usepackage{amsmath}
\usepackage{float}
\usepackage{epstopdf}
\usepackage{url}

\begin{document}


\thispagestyle{empty} \vspace*{0.8cm}\hbox
to\textwidth{\vbox{\hfill\huge\sf Commun. Theor. Phys.\hfill}}
\par\noindent\rule[3mm]{\textwidth}{0.2pt}\hspace*{-\textwidth}\noindent
\rule[2.5mm]{\textwidth}{0.2pt}


\begin{center}
\LARGE\bf Rayleigh-Taylor instability under multi-mode perturbation: discrete Boltzmann modeling with tracers$^{*}$
\end{center}

\footnotetext{\hspace*{-.45cm}\footnotesize $^\dag$Corresponding author, E-mail: xu\_aiguo@iapcm.ac.cn }

\begin{center}
\rm Hanwei Li$^{\rm a)}$, \ \ Aiguo Xu$^{\rm a,b,c)\dagger}$, \ \ Ge Zhang$^{\rm d)}$,\ and \ Yiming Shan$^{\rm a)}$
\end{center}

\begin{center}
\begin{footnotesize} \sl
\textit{}{Laboroatory of Computational Physics, Institute of Applied Physics and Computational Mathematics, Beijing 100088, China}$^{\rm a)}$ \\
\textit{Center for Applied Physics and Technology, MOE Key Center for High Energy Density Physics Simulations, College of Engineering, Peking University, Beijing 100871, China}$^{\rm b)}$ \\
\textit{State Key Laboratory of Explosion Science and Technology, Beijing Institute of Technology, Beijing 100081, China}$^{\rm c)}$ \\
\textit{Energy Resources Engineering Department, Stanford University, 367 Panama St, Stanford, CA 94305-2220, USA}$^{\rm d)}$ \\
\end{footnotesize}
\end{center}

\begin{center}
\footnotesize (Received XXXX; revised manuscript received XXXX)

\end{center}

\vspace*{2mm}

\begin{center}
\begin{minipage}{15.5cm}
\parindent 20pt\footnotesize
The Rayleigh-Taylor Instability (RTI) under multi-mode perturbation in compressible flow is probed via the Discrete Boltzmann Modeling (DBM) with tracers. The distribution of tracers provides clear boundaries between light and heavy fluids in the position space. Besides, the position-velocity phase space offers a new perspective for understanding the flow behavior of RTI with intuitive geometrical correspondence. The effects of viscosity, acceleration, compressibility, and Atwood number on the mixing of material and momentum and the mean non-equilibrium strength at the interfaces are investigated separately based on both the mixedness defined by the tracers and the non-equilibrium strength defined by the DBM. The mixedness increases with viscosity during early stage but decreases with viscosity at the later stage. Acceleration, compressibility, and Atwood number show enhancement effects on mixing based on different mechanisms. After the system relaxes from the initial state, the mean non-equilibrium strength at the interfaces presents an initially increasing and then declining trend, which is jointly determined by the interface length and the macroscopic physical quantity gradient. We conclude that the four factors investigated all significantly affect early evolution behavior of RTI system, such as the competition between interface length and macroscopic physical quantity gradient. The results contribute to the understanding of the multi-mode RTI evolutionary mechanism and the accompanied kinetic effects.
\end{minipage}
\end{center}

\begin{center}
\begin{minipage}{15.5cm}
\begin{minipage}[t]{2.3cm}{\bf Keywords:}\end{minipage}
\begin{minipage}[t]{13.1cm}
Rayleigh-Taylor instability; multi-mode perturbation; Discrete Boltzmann modeling; tracers; non-equilibrium effects; kinetic modeling
\end{minipage}\par\vglue8pt

\end{minipage}
\end{center}

\section{Introduction}\label{sec:level1}
Rayleigh-Taylor instability (RTI) occurs at the interface where the heavy fluid is supported or accelerated by the light fluid$^{[1,2]}$, and the perturbation does not disappear if the interface is perturbed, but grows with time. RTI is widespread in nature as well as in industry and is an important phenomenon in many fields of science and engineering, such as the filling of freshwater layers by seawater$^{[3]}$, the outburst of supernovae in outer space$^{[4]}$, two-component Bose-Einstein condensation$^{[5]}$, and microsecond and millisecond pulsed discharges in water$^{[6]}$, etc. In particular, in the inertial confinement fusion (ICF), it is necessary to suppress the development of RTI to achieve the conditions required for ignition$^{[7-12]}$. Therefore, an in-depth study of RTI is of great importance in both basic science and engineering applications.

Realistic RTI is generally caused by multi-mode perturbations. However, from the perspective of theoretical research, the problems are often divided into single-mode and multi-mode according to the initial perturbation of the interface. The knowledge gained from single-mode research provides a mechanism basis for the study of multi-mode perturbation$^{[13]}$. Early studies of RTI focused on the single-mode phenomenon for the convenience of theoretical analysis. This paper focuses on the multi-mode RTI and explores the development law and inner mechanism of complex flows. The research methods of RTI can be generally divided into three categories, experiment, theory, and numerical simulation. Sharp$^{[14]}$ made a good summary of the early research of RTI and divided the development of instability into four stages, linear growth, weakly nonlinear growth, strongly nonlinear stage and chaotic mixture stage. It is very difficult to solve the strongly nonlinear stage and chaotic mixture stage analytically, which leads to the relatively limited results obtained from the theoretical analysis, and the mechanism of the RTI nonlinear development stage is still far from clear$^{[15]}$. The results obtained from experiments are intuitive. Ramaprabhu et al.$^{[16]}$ studied the self-similar evolution to turbulence of a multi-mode RT mixing at small density differences by particle image velocimetry (PIV) and high-resolution thermocouple measurement, and presented first-, second-, and third-order statistics with spectra of velocity and temperature fields. Analysis of the measurements has shed light on the structure of mixing as it develops to a self-similar regime in this flow. Mueschke, et al.$^{[17]}$ used a series of experimental methods to quantify the initial multi-mode interfacial velocity and density perturbations present at the onset of a small Atwood number (A), incompressible, miscible, RTI-driven mixing layer. The results show that the measured initial conditions describe an initial anisotropic state in which the streamwise and spreading perturbations are independent of each other. The early evolution of the density and vertical velocity variance spectra suggest that velocity fluctuations are the dominant mechanism driving the development of instability. Morgan, et al.$^{[18]}$ performed experiments to observe the growth of the turbulent Rayleigh-Taylor unstable mixing layer generated between air and SF$_6$. They used the membraneless vertical oscillation technique to make the three-dimensional multi-mode perturbations on the diffuse interface, and 20 experiments were performed to establish a statistical ensemble. The average perturbation from the ensemble was found and used as input for a numerical simulation, and they observed good qualitative agreement between the experiment and simulation. However, the disadvantages of this approach are also evident due to the high cost of the experiment, the relatively limited scope, and the considerable dependence of the results on observation means.

Because both experimental and theoretical studies have their own limitations, numerical simulation methods have been used extensively over the years. Ramaprabhu, et al.$^{[19]}$ investigated the effect of initial conditions on the growth rate of turbulent RT mixing by using the monotone integrated large-eddy simulation (MILES) to solve the three-dimensional incompressible Euler equation with numerical dissipation. Dimonte, et al.$^{[20]}$ performed experimental and numerical simulations on the RTI with a complex acceleration history g(t) and it was found that the dominant bubbles and peaks growing in the initial unsteady phase were shredded by the trailing structures during the stable deceleration phase, which reduced their diameter and increased the mixing of the liquid. Youngs, et al.$^{[21]}$ summarized the study on self-similar mixing caused by RTI and described a series of high resolution large eddy simulations. They investigated the properties of high Reynolds number self-similar RT mixing, and provided data for the calibration of engineering models. Zhang, et al.$^{[22]}$numerically studied the self-similar nonlinear evolution of the multi-mode ablative Rayleigh-Taylor instability (ARTI) in both two and three dimensions. The results showed that ablation-driven vorticity accelerates the bubble velocity and prevents the transition from the bubble competition to the bubble merger regime at large initial amplitudes. Liang, et al.$^{[23]}$ performed high-resolution direct numerical simulations of multi-mode immiscible RTI with a low Atwood number using the modified phase field lattice Boltzmann method. The effects of Reynolds numbers on the evolving interface dynamics and bubble/spike amplitude were investigated. The numerical results showed that, when the Reynolds number is large enough, the immiscible RTI can be divided into linear growth, saturated velocity growth, and chaotic development stages. The RTI induces a complex interface topology at the later stage. When the Reynolds number is reduced to small ones, the multiple perturbations of the RTI merge into larger perturbations at the initial stage. Yilmaz, et al.$^{[24]}$ used the large eddy simulation to analyze three-dimensional, multi-mode RTI at high Atwood number. The results showed that RTI at high Atwood number develops rapidly due to the increasing growth rate and velocity of spikes, increasing asymmetry in the mixing region, and more intensive interactions in the nonlinear phase. It was also found that the interaction of the spike fronts with their surroundings is the main mechanism of turbulence generation and transition to the turbulent phase. Livescu, et al.$^{[25]}$ gave direct numerical model results of three dimensional multi-mode RTI at Atwood number up to 0.9, after a large development of the layer width, and they ran additional branching simulations in reverse and zero gravity conditions. The effects of acceleration variation on mixed-layer structure and turbulence are highlighted. Hamzehloo, et al.$^{[26]}$ performed direct numerical simulations of two- and three-dimensional, single- and multi-mode incompressible unmixed-phase RTI using a phase-field approach and a high-order finite-difference scheme. Different combinations of Atwood number, Reynolds number, surface tension, and initial perturbation amplitude were investigated. The results show that three-dimensional multi-mode RTI initially exhibits an exponential growth rate similar to that of single-mode. Ding, et al.$^{[27]}$ investigated the RTI at single- and dual-mode interfaces under strong accelerations by molecular dynamics simulations. They found that the growth behavior of microscopic RTI and the underlying mechanisms show  considerable differences from macroscopic RTI. The microscopic RTI exhibits much weaker nonlinear effect. Wang, et al.$^{[28]}$ investigated the effect of interfacial diffusion on turbulent transition in sparse-driven flows by implicit large eddy simulations with three dimensional multi-mode perturbations applied to diffusion interfaces of different thicknesses.

These aforementioned works have achieved some good results in the field of multi-mode RTI research, but since they are generally based on the traditional set of macroscopic hydrodynamic equations (e.g., Euler or Navier-Stokes equations), all of them seldom focus on the rich and complex non-equilibrium effects of fluid systems during RTI development. The Discrete Boltzmann Modeling (DBM) has been applied to the study of various fluid instability problems as an effective physical modeling and simulation method$^{[29-38]}$, and the DBM focuses on the interaction of different non-equilibrium behaviors in complex flows. Lai, et al.$^{[29,30]}$ first implemented the modeling and simulation of RTI systems using a single-relaxation DBM. They found that the compressibility effect has a two-stage effect on the RTI interface evolution. The compressibility inhibits RTI development in the early stage but accelerates it in the later stage. Chen, et al.$^{[31,32]}$ studied the correlation between the inhomogeneities of macroscopic quantities such as density, temperature, flow velocity, pressure and the non-equilibrium strength during RTI evolution and the effects of viscosity, heat conduction and Planter number on the growth of interfacial perturbations and on the above correlations by linking both macroscopic quantities and non-equilibrium features based on multi-relaxation time DBM. Based on the above studies, Chen, et al.$^{[33,34]}$ further investigated the competitive cooperation between Richtmyer-Meshkov instability (RMI) and RTI based on the multi-relaxation DBM. They introduced also the morphological boundary length to study the competitive mechanism of the coupled RT-KHI system. Lin, et al.$^{[35]}$ studied the RTI in a two-component compressible fluid for high and low Reynolds numbers. Ye, et al.$^{[36]}$ continued to study the effect of Knudsen number on RTI in compressible fluids by DBM, and found that the higher Knudsen number has a stronger suppression effect on the RTI, but has a stronger enhancing effect on the hydrodynamic non-equilibrium (HNE) and thermodynamic non-equilibrium (TNE) effects. Chen, et al.$^{[37]}$ studied the specific heat ratio effect of RTI in compressible fluids, and discussed the variation of non-equilibrium strength with specific heat ratio. By introducing tracers in DBM, Zhang, et al.$^{[38]}$ provided fine structure images with clear interfaces in RTI evolution, defined the mixing degree based on tracers, revealed the mixing process of light and heavy fluids, and further investigated the effect of compressibility and viscosity on RTI.

In recent years, the DBM-based RTI studies focused on the non-equilibrium behavior characteristics which are missed by the Navier-Stokes (N-S) model and have yielded fruitful results. But these studies are basically based on single-mode perturbation. In this paper, we study the complex flow and mixing of multi-mode perturbation RTI based on the coupled modeling of DBM and tracers proposed by Zhang, et al.$^{[38]}$. The tracers provide a Lagrangian view of the same flow evolution. This paper is organized as follows. The numerical simulation methods and physical modeling are presented in Section 2. The Numerical results and discussion are presented in Section 3. Finally, conclusions are made in Section 4.

\section{Model and method}\label{sec:method}
Traditional fluid models such as the N-S are based on the continuum assumption and near-equilibrium approximation. The continuous assumption brings an intrinsic physical constraint that the characteristic length of flow behaviors is large enough compared to the mean free path of molecules. In that case, the system is always at thermodynamic equilibrium state or near equilibrium state. However, there are always various spatio-temporal scales in some compelx flow systems. The large structure and slow-varying behaviors can basically be well described with the help of traditional fluid models. However, as the scale of the structure of interest gradually decreases, the relative molecular spacing is gradually no longer a negligibly small quantity. As the kinetic behavior of interest becomes more fast, the system has less sufficient time to relax to the thermodynamic equilibrium, and consequently the degree of TNE gradually increases. It can be seen that the physical rationality of N-S begins to be challenged as the spatio-temporal scale of the behavior of interest decreases. At the same time, these behaviors often occur at the spatio-temporal scales that are not available or easily accessible to microscopic molecular dynamics simulations. These spatio-temporal behaviors that challenge both macroscopic continuum and microscopic molecular dynamics models are far from being adequately studied because of the inadequacy of description methods

\subsection{The construction of DBM} \label{DBM_cons}
Discrete Boltzmann modeling is a kinetic modeling method that has been rapidly developed in recent years for the above mesoscale behavior studies. The establishment of the DBM model generally requires three steps: (i) to linearize the collision operator of revised Boltzmann equation, (ii) to discretize the particle velocity space, and (iii) to describe the non-equilibrium state and pick non-equilibrium information. Of these, the first two steps are coarse-grained physical modeling, which requires that the system behavior of interest cannot be changed by model simplification; the third step is the purpose and core of the DBM. Based on the following realities: (1) the intermolecular potential may be far from being as weak and simple as required by the Boltzmann equation, (2) the degree of non-equilibrium of interest may far exceed the quasi-equilibrium required by the original Bhatnagar-Gross-Krook (BGK) model$^{[39]}$, and the study of the kinetic behavior of most systems cannot be carried out solely on the basis of the original BGK kinetic theory, the actual BGK model used is in fact a modified version based on the mean-field theory. The duty of mean-field theory is twofold: (1) to supplement the description of the intermolecular potential effects missed by Boltzmann's equation, and (2) to modify the applicability of BGK so that it can be extended to higher levels of non-equilibrium.

In a review article of 2012, Xu, et al.$^{[40]}$ pointed out that, under the framework of discrete Boltzmann equation and under the condition that do not use non-physical Boltzmann equation and kinetic moments, the non-conservative moments of $(f-f^{eq})$ can be used to check and describe how and how much the system deviates from the thermodynamic equilibrium, and to check and describe corresponding effects due to deviating from the thermodynamic equilibrium. This was starting point for the current DBM approach,  where $f$ and $f^{eq}$ are the distribution function and the corresponding equilibrium state distribution function. In a research article of 2015, Xu, et al. $^{[41]}$ proposed to open a phase space based on the non-conservative moments of $(f-f^{eq})$ and to describe the depth of TNE using the distance between a state point to the origin in the phase space or its subspace. In the RGD31 conference held in 2018, Xu, et al.$^{[42]}$ further developed the non-conservative moment phase space description methodology. They proposed to use the distance $D$ between two state points to roughly describe the difference of the two states deviating from their thermodynamic equilibriums from a perspective, and the reciprocal of distance, $1/D$, is defined as a similarity of deviating from thermodynamic equilibrium. The mean distance $\bar{D}$ during a time interval is used to roughly describe the difference of the two corresponding kinetic processes from a perspective, and the reciprocal of $\bar{D}$, $1/\bar{D}$, is defined as a process similarity. In a review article of 2021, Xu, et al.$^{[43]}$ extended the phase space description methodology to any system characteristics as follows: Use a set of (independent) characteristic quantities to open phase space, and use this space and its subspaces to describe the system properties. A point in the phase space corresponds to a set of characteristic behavior of the system.  Distance concepts in the phase space or its subspaces are used to describe the difference and similarity of behavior. Therefore, DBM is a further development of the statistical physics phase space description method in the framework of the discrete Boltzmann equation, which presents an intuitive geometric correspondence for the complex non-equilibrium behavior. DBM can surpass the traditional fluid modeling in terms of both depth and width of non-equilibrium behavior description$^{[43-45]}$

Simulating RTI systems with DBM requires consideration of acceleration effects. Lai, et al.$^{[29]}$ proposed a DBM with acceleration. The evolution equation can be written as follow,
\begin{equation}
\frac{\partial f}{\partial t} + \mathbf{v} \cdot \frac{\partial f}{\partial {\mathbf{r}}} - \frac{\mathbf{a}\cdot \left ( \mathbf{v} - \mathbf{u} \right )}{RT}f^{eq}=
- \frac{1}{\tau} \left ( f - f^{eq} \right )
\label{BGK-BE}
\end{equation}
where $f=f(\textbf{r}, \textbf{v}, t)$ is the distribution function in phase space. $\textbf{r}$ indicates spatial position, $\textbf{v}$ is the macro flow velocity, $t$ is the time, $\textbf{a}$ corresponds to external body force, $R$ is the gas constant, $T$ is the temperature, $\tau$ is the relaxation time,and $f^{eq}$ is the local equilibrium state distribution function.

The discrete velocity model with two-dimensional sixteen velocities (D2V16) is used to discretize the continuous velocity space, and the corresponding set of macroscopic hydrodynamic equations can be obtained by Chapman-Enskog (C-E) multiscale expansion. The DBM retaining the seven kinetic moments can completely cover the two-dimensional compressible N-S set of equations in terms of the descriptive function of fluid dynamics, and according to the existing research results, the DBM provides reliable simulation results of the flow field compared with the traditional methods of solving the N-S set of equations$^{[41,46]}$. Among these seven kinetic moments, only the first three (density moment, momentum moment, and energy moment) are conserved moments that remain constant as the system tends to or leaves equilibrium, while the rest of the kinetic moments are non-conserved.

The description and extraction of non-equilibrium information of the flow field is the purpose and core of DBM modeling, and the non-equilibrium behavior characteristics can be described in detail by the non-conservative kinetic moment of $\left(f-f^{eq}\right)$, and the non-equilibrium characteristic quantity $\boldsymbol{\Delta}^{*}_{m, n}$ can be defined accordingly,
\begin{equation}
\boldsymbol{\Delta}_{m,n}^{*}=\mathbf{M}_{m,n}^{*}\left ( f \right ) - \mathbf{M}_{m,n}^{*}\left ( f^{eq} \right )
\label{eq02}
\end{equation}
where $\mathbf{M}_{m,n}^{*} \left( f \right)$ describes the kinetic moment of the distribution function $f$ with respect to the particle thermal fluctuation velocity $\mathbf{v}^* = \mathbf{v} - \mathbf{u}$, i.e., the kinematic central moment, $m$ denotes the number of velocities used in the moments and $n$ denotes the tensor order.The non-equilibrium eigenmoments $\boldsymbol{\Delta}_{m,n}^{*}$ describe the TNE that characterize the purely thermal rise and fall of microscopic particles after deducting the macroscopic flow. These non-conservative moments are interrelated and together they constitute a relatively complete description of the non-equilibrium state and behavior of the system. The $\boldsymbol{\Delta}_{2}^{*}$ and $\boldsymbol{\Delta}_{3,1}^{*}$ denote the non-equilibrium eigenmoments defined earlier, which correspond to the viscous stress and heat flow terms, respectively, in the set of fluid dynamics equations called non-organized momentum flux (NOMF) and non-organized energy flux (NOEF). The $\boldsymbol{\Delta}_{3}^{*}$ and $\boldsymbol{\Delta}_{4,2}^{*}$ represent higher-order non-equilibrium effects beyond the N-S level, representing the viscous stress flux and heat flow flux.

\subsection{Tracers coupled in DBM} \label{TRACERS}
Tracers describing fluid flow belong to a special class of particles. A particle immersed in a fluid always interacts with the surrounding fluid and the kinetic state of the particle can be characterized by the Stokes number (Stk),
\begin{equation}
Stk = \frac{u_0 \cdot \tau_p}{l_0}
\label{eq_stk}
\end{equation}
where $u_0$ is the local flow velocity, $\tau_p$ is the characteristic relaxation time of the particle, and $l_0$ denotes the characteristic length (generally the diameter of the particle is chosen). The larger $Stk$ indicates that the inertial effect of the particle is larger, and the particle is easy to break away from the local flow when the flow velocity changes drastically; when $Stk$ is much smaller than 1, the particle will follow the flowline motion closely, and in this case, the effect of the immersed particle on the surrounding fluid is not significant. By adjusting the relaxation time $\tau_p$ of the particles, $Stk$ can be set to a small enough magnitude that the immersed particles can be used as tracers in the flow field.

The velocity of a tracer is determined by a combination of its location and the local local flow velocity
\begin{equation}
\mathbf{u}_p \left (\mathbf{r}_k \right )=\int_{D} \mathbf{u} \left ( \mathbf{r} \right )
\cdot \delta \left ( \left | \mathbf{r} - \mathbf{r}_k \right | \right ) dD
\label{eq_velocity_tracer}
\end{equation}
where $\mathbf{u}_p$ is the velocity of the particle's motion, $\mathbf{r}_k$ denotes the position where the $k$-th particle is located, and $\delta$ is the Dirac function chosen to obtain the velocity of the tracer particle from the surrounding fluid$^{[47]}$. The above equation applies to the entire fluid domain $D$.

After the flow field is discretized in the numerical simulation, the point particles will basically fall between the grid nodes, so the continuous Dirac function needs to be approximated by discretizing,
\begin{equation}
\mathbf{u}^t \left( \mathbf{r}_k \right )= \sum_{i,j} \mathbf{u}_{i,j}^{t} \psi \left( \mathbf{r}_{i,j},\mathbf{r}_k \right )
\label{eq_velocity_discretize}
\end{equation}
where $\psi$ is the discrete form of the Dirac function $\delta$. In the two-dimensional case, $\psi$ can be decomposed as the product of two one-dimensional discrete Dirac functions,
\begin{equation}
\psi \left( \mathbf{r}_{i,j},\mathbf{r}_k \right )=\psi \left( \left | \mathbf{r}_{i,j} - \mathbf{r}_k \right | \right)=\varphi \left ( \Delta r_x  \right ) \cdot \varphi \left ( \Delta r_y  \right )
\label{eq_dirac_func}
\end{equation}

The weighting function $\varphi$ applied here is as follow,
\begin{equation}
\varphi \left ( \Delta r_x  \right )=\left \{ \begin{array} {ll}
\left \{ 1+\textup{cos}\left [ \left ( \Delta r_x/\Delta x \right ) \cdot \pi/2 \right ] \right \},  & \Delta r_x \leq 2\Delta x \\
0 ,& \Delta r_x >  2\Delta x \end{array} \right.
\label{eq_varphifunc}
\end{equation}

Once the velocity is determined, the trajectory of the each tracer can be captured by the equation of motion.
\begin{equation}
\frac{\partial\mathbf{r}_k}{\partial t}=\mathbf{u}_p
\left ( \mathbf{r}_k \right )
\label{eq_motion}
\end{equation}
In order to update the position information of the tracers, the equations of motion of the tracers are discretized using a 4th-order Runge-Kutta scheme.

\subsection{Initial settings} \label{Initial settings}

Based on the above model and method, this paper uses a single medium fluid to simulate the RTI flow caused by multi-mode perturbation in a compressible non-isothermal case. The fluid system consists of two parts with different temperatures, the initial temperature of the upper part of the fluid is $T_u$ and the initial temperature of the lower part of the fluid is $T_b$, the interface is located in the middle of the fluid, the left and right boundaries are set as periodic boundaries, and the upper and lower boundaries are set as solid wall boundaries. In order to induce RTI to occur and destroy the unstable equilibrium state of the flow field, an initial perturbation is added to the fluid interface at the initial moment$^{[30]}$.
\begin{equation}
y_c(x)= \sum_{n=21}^{30}[a_n\cos(k_nx)+b_n\sin(k_nx)]
\label{eq_perturbation}
\end{equation}
$k_n=2\pi n/L_x(L_x=NX\cdot dx=2d)$. $a_n$ and $b_n$ are random numbers uniformly distributed between $0-1$. The density distribution of the fluid satisfies the hydrostatic equilibrium condition:
\begin{equation}
\frac{\partial p(y)}{\partial y}=-g\rho_0(y)
\label{eq_pressure}
\end{equation}
The initial conditions of the flow field are as follows$^{[29,38]}$,
\begin{equation}
\left\{
\begin{split}
T_0(y)&=T_u\\
\rho_0(y)&=\frac{p_0}{T_u}\exp [\frac{g}{T_u}(d-y)],y>y_c(x)\\
T_0(y)&=T_u\\
\rho_0(y)&=\frac{p_0}{T_b}\exp [\frac{g}{T_u}(d-y_c(x))-\frac{g}{T_b}(y-y_c(x))],y<y_c(x)
\end{split}
\right.
\label{eq_initial_condition}
\end{equation}
where $p_0$ is the initial pressure at the top of the upper half of the fluid. The pressure at the interface between the top half of the fluid and the bottom half of the fluid is equal,
\begin{equation}
p_c=\rho_u T_u=\rho_b T_b
\label{eq_interface_pressure}
\end{equation}
where $\rho_u$ and $\rho_b$ are the densities of the upper and lower parts of the fluid on each side of the interface, respectively. The initial Atwood number at the interface is defined as follows.
\begin{equation}
A=\frac{\rho_u-\rho_b}{\rho_u+\rho_b}=\frac{T_b-T_u}{T_b+T_u}
\label{eq_At}
\end{equation}

The other parameters required in the simulation are shown in Table 1, and they are all dimensionless quantities. In the initial setup of the tracers, a total of about 250,000 tracers are set up in order to ensure sufficient resolution under the condition of saving computational costs. These tracers are randomly and uniformly distributed over the entire simulation area. The tracers locate above the fluid interface in the initial conditions are labeled as \textit{type-a}; those locate below the fluid interface are are labeled as \textit{type-b}, which makes it possible to distinguish the distribution and origin of the tracers during the simulation. An additional 256 tracers, labeled \textit{type-c}, are set at the interface, and these tracers are able to record the interface deformation as well as the changes in the physical quantities at the interface.
\begin{center}
	\tabcolsep=15pt
	\small
	\renewcommand\arraystretch{1.2}
	\begin{minipage}{15.5cm}{
			\small{\bf Table 1.} Initial physical parameters for RTI simulation }
	\end{minipage}
	\vglue5pt
	\begin{tabular}{lccc}
	$Parameters$                      &   $Values$   \\[3pt]
	Grid number ($NX=NY$)            &   512$\times$512\\
	Time step ($dt$)                 &   2.0$\times$10$^{-5}$ \\
	Grid size ($dx$=$dy$)            &   1.0$\times$10$^{-3}$ \\
	Relaxation time ($\tau$)         &   3.0$\times$10$^{-5}$ \\
	Acceleration ($g$)               &   1.0 \\
	Upper fluid temperature ($T_u$)  &   1.0 \\
	Top pressure ($p_0$)       &   1.0 \\
	Atwood number ($A$)             &   0.6 \\
	\end{tabular}
\end{center}

\section{Results and discussion}\label{sec:results}

\subsection{Multi-view RTI Evolution}\label{subsec:evolution}
\begin{figure}
	\centerline{\includegraphics[width=12cm]{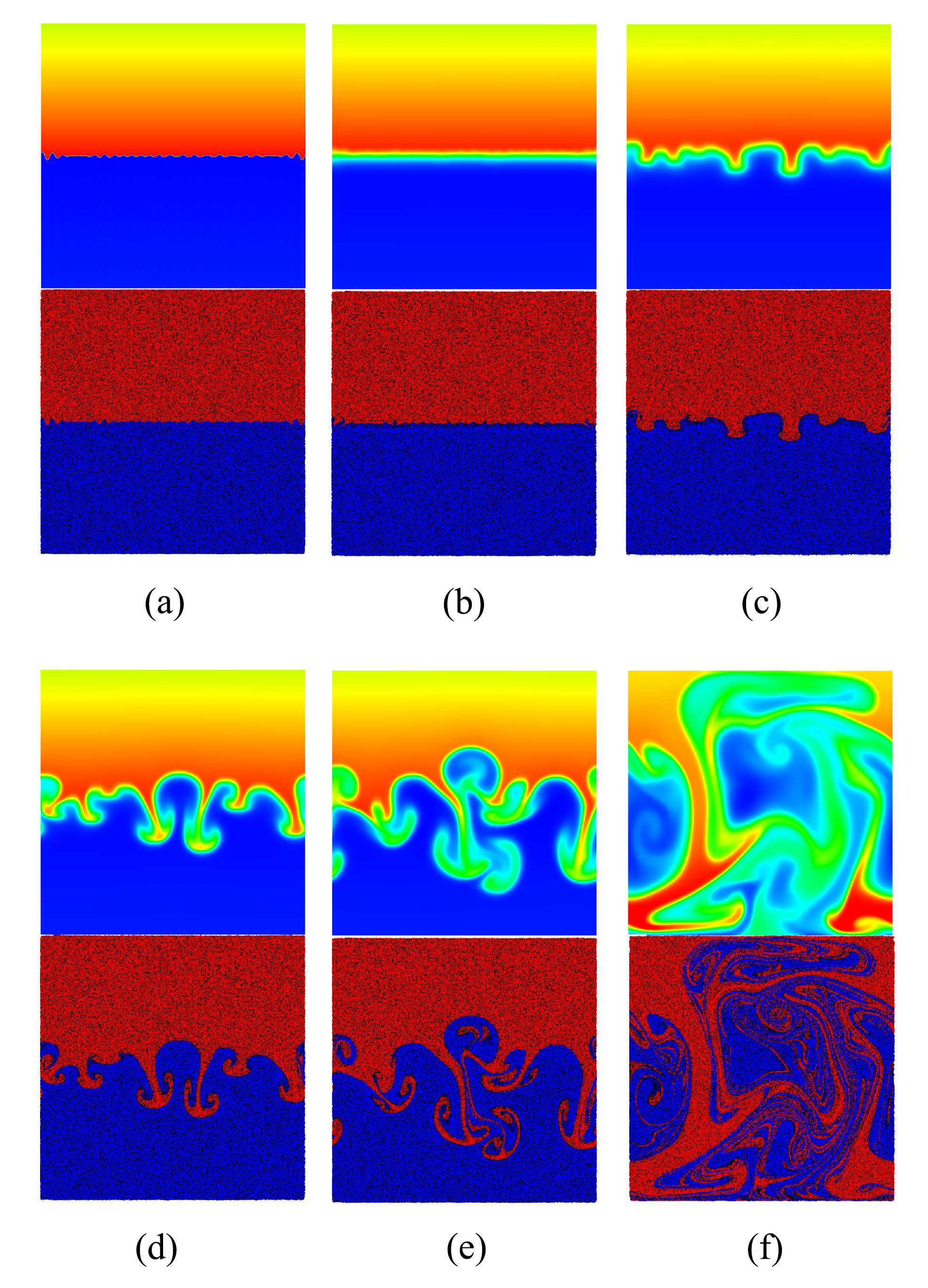}}
	\begin{center}
		\parbox{15.5cm}{\small{\bf Fig.1.}  Comparison between density contours and tracer distribution patterns at different times; the upper half in each part shows the density contour and the lower half in each part shows tracer distribution. (a) $t=0$; (b) $t= 0.5$; (c) $t=1.5$; (d) $t=2.0$; (e) $t=2.5$; (f) $t=4.5$. }
	\end{center}
\end{figure}	

Figure 1 shows a comparison between the density contours and the patterns of the tracers during the mixing of light and heavy fluids caused by multi-mode perturbation RTI. In this paper, a single-fluid model is used to study the mixing caused by multi-mode perturbations in compressible and  miscible fluid systems, which is closer to the actual conditions in nature and engineering, but the density transition layer is formed near the fluid interface due to the mixing of light and heavy fluids. As the flow field evolves, the interface becomes more complex and it becomes very difficult to obtain the fine structure of the flow field interface by density contour. However, by labeling two types of tracers (\textit{type-a} and \textit{type-b}) with different colors, the tracer distribution patterns can still provide a clear boundary of two fluids at each moment of RTI evolution even though the flow field evolution caused by multi-mode perturbation is more complex than that caused by single-mode perturbation. In Figure 1 (a)-(d), the density contours (on the upper half) and the tracer distribution patterns (on the lower half) are almost identical, which can prove the validity of the tracers. With further mixing of the flow field, the advantages of the tracer description gradually emerge. At the later stage of fluids mixing, it is very difficult to distinguish the components somewhere in the flow field using the density contours. From Figure 1 (e)-(f), the tracer distribution patterns always shows a clear interface structure, even if there are a large number of separated and broken fluid areas in the flow field. Compared with single-mode perturbation$^{[38]}$, the fluids mixing caused by multi-mode perturbation contains more abundant small structures and the interface shape and evolution are more complex, so the advantage of tracers to describe the fine structure is more fully reflected. Inevitably, there are some white noise spots on patterns of the tracers, and the resolution can be significantly improved by increasing the number of tracers. The number of tracers depends on the trade-off between resolution and computational cost.
\vspace*{4mm}

\subsection{Tracers phase space}\label{subsec:phase_space}
Each tracer carries the position and velocity information of a point in the flow field. Zhang, et al.$^{[38]}$ used the velocity information carried by the tracers to give the velocity phase diagrams of two particles at different moments in the case of single-mode perturbation RTI. They separated the analysis of heavy and light fluids, which is important for understanding RTI. The concept of phase space is commonly used in statistical physics to describe all possible microscopic motion states of a system, such as the $\mu$-space and $\Gamma$-space. $\mu$-space is the four-dimensional space containing $(x,y,ux,uy)$ under two-dimensional conditions. Since it is not easy to display the space above three dimensions, we use three-dimensional subspaces of $\mu$-space, such as $(x,ux,uy)$ and $(y,ux,uy)$ to describe the motion of tracers at different moments. The pattern of tracers distribution and the velocity phase space used by Zhang et al. are essentially two-dimensional subspaces of $\mu$ space. However, the two-dimensional phase space describes the position information separately from the velocity information, and cannot give the distribution relationship between the velocity and the position at a certain moment, where the tracer phase space description method has a new breakthrough. Figure 2 gives the three-dimensional subspace descriptions of two kinds of particles for $(1)(2)(3)(4)$ at four moments, corresponding to heavy fluids$((a),(b))$ and light fluids$((c),(d))$. Where $((a),(c))$ represents the $(x,ux,uy)$ subspace and $((b),(d))$ represents the $(y,ux,uy)$ subspace.

At $t = 0.5$, RTI occurs initially and most of the tracers are still at the origin. At the same time, diffusion occurs between the two fluids due to the density gradient, causing the tracers near the interface to acquire a certain velocity. As shown in the $x-y$ projection of Figure 2 $a(1)$, the $ux$ of the heavy fluid particles \textit{type-a} fluctuates around $ux = 0$, which is related to the initial perturbation. The $uy$ of \textit{type-a} particles shows a band distribution along the $x$ direction, and the main part still fluctuates around $uy=0$. However, because the heavy fluid particles near the  interface diffuse downward, a not narrow band structure is formed in the region of $uy<0$. As shown in Figure 2 $a(2)$, the $ux$ of \textit{type-a} particles has a symmetric spike distribution along the $y$ direction, and the larger $y$ is, the narrower the velocity distribution interval is. After $y$ is larger than a certain value $ux$ is about 0. And the $uy$ also has a similar spike distribution along the $y$ direction, except that there is an abrupt change near $y=0.2$. This indicates that the perturbations at the interface are not massively transmitted to other locations in the fluid system at this time. At $t = 1.5$, both \textit{type-a} particles and \textit{type-b} particles show a very regular and beautiful characteristic pattern in either subspace. The velocity phase diagrams ($y-z$ projection in the second row) of both form a laminar "shuttle" structure. The tip of the velocity diagram of the \textit{type-a} particles is facing downward, while the tip of the \textit{type-b} particles is facing upward, indicating that most of the two particles have opposite trends of motion. The $ux$ and $uy$ form a continuous undulating "crest-valley" structure along the $x$ direction respectively, which corresponds to the final stage of linear growth of RTI. And $ux$ and $uy$ still have symmetric spike distribution pattern along the $y$ direction respectively, only that the lower part of the distribution is fuller at this time, making the characteristic $y$ coordinate with zero velocity become larger, which indicates that the mixing zone is obviously wider and the spike inserts the bubble deeper. As the mixing intensifies, the pattern in the three dimensional subspace starts to become more turbulent. At $t = 2.5$, the spike of $ux$ and $uy$ along the $y$ direction can be roughly observed, and the mixing zone is getting closer to the boundary. At $t = 4.5$, the flow field has been initially mixed. Some interesting observations in the phase space of tracers are presented here, and the study of their dynamical processes is still important for further understanding.
\vspace*{4mm}
\begin{figure}
	\centerline{\includegraphics[width=16cm]{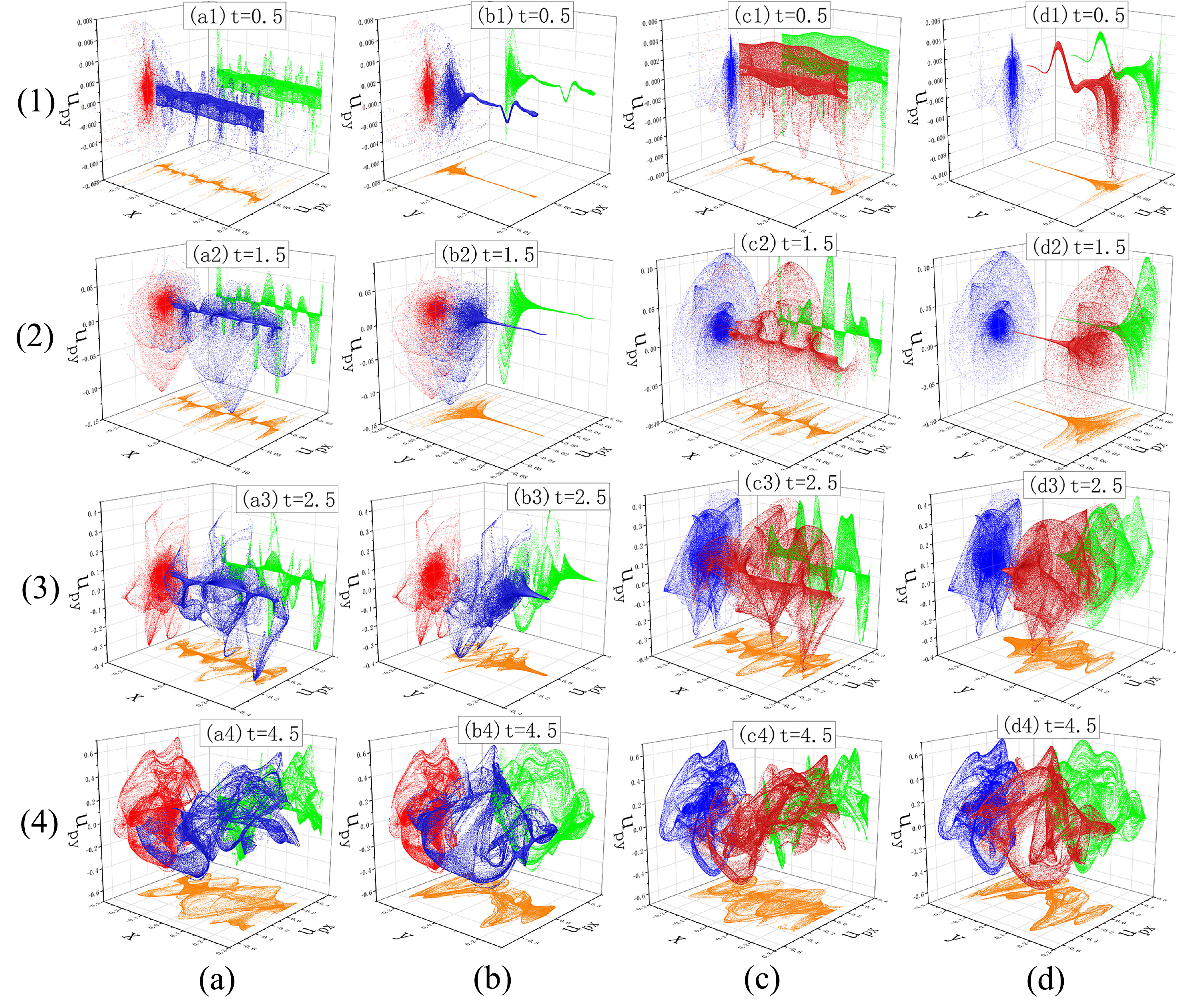}}	
	\begin{center}
		\parbox{15.5cm}{\small{\bf Fig.2.}  Phase space description of two kinds of tracers at different times: select $(1)(2)(3)(4)$ four time display $(t =0.5,1.5,2.5,4.5)$; $(a)(b)$ is the $(x,ux,uy)$ and $(y,ux,uy)$ subspace of heavy fluid particles, $(c)(d)$ is the $(x,ux,uy)$ and $(y,ux,uy)$ subspace of light fluid particles. }
	\end{center}
\end{figure}

\subsection{Non-equilibrium strength}\label{subsec:Non-equilibrium}
The interface is the main place for the occurrence and development of RTI, so it is necessary to pay extra attention to study it in depth. In particular, the variation of physical quantities at the interface is important for the control of the interface instability in ICF that we are interested in. Transport theory tells us that non-equilibrium is the fundamental driver of system evolution. For fluid systems, the thermodynamic non-equilibrium intrinsic to the system drives the evolution of the flow field toward convergence to equilibrium. The non-equilibrium behavior is extremely complex, while the degree of non-equilibrium description is diverse. As mentioned in the previous section, the non-equilibrium behavior can be characterized by the nonconservative kinetic moments of $(f-f^{eq})$. The various non-equilibrium descriptions are inherently complementary and related. The DBM is capable of both describing specific non-equilibrium states from their own perspectives using the independent TNE components, and a high-dimensional phase space that can be tensed using the individual TNE components. In the phase space, the origin corresponds to a thermodynamic equilibrium state, and any point in the phase space corresponds to a specific thermodynamic non-equilibrium state, and the distance between the two can be defined as the non-equilibrium strength.

The TNE information of the tracers can be extracted by interpolating the TNE information on the grid points in the flow field, and thus the following can be calculated,
\begin{equation}
\overline{\boldsymbol{\Delta}}_{m,n}^{*}=\frac {1}{\Lambda_{i}} \int_{\Lambda_{i}} \boldsymbol{\Delta}_{m,n}^{*} \left (x,y \right ) dl \approx \frac
{\sum_{1}^{N} \boldsymbol{\Delta}_{m,n,k}^{*} }{N}
\label{eq:TNE}
\end{equation}
where $\Lambda_{i}$ indicates the length of interface and $N$ is the total number of \textit{type-c} tracers.The mean non-equilibrium strength of the system at the interface is defined as follows
\begin{equation}
\overline{D}\approx \frac{\sum_{1}^{N} D_k}{N}
=\frac{1}{N}\sum_{1}^{N}\sqrt{ \left( \boldsymbol{\Delta}_{2,k}^{*} \right)^2 + \left ( \boldsymbol{\Delta}_{3,k}^{*} \right)^2 + \left( \boldsymbol{\Delta}_{3,1,k}^{*} \right)^2 + \left( \boldsymbol{\Delta}_{4,2,k}^{*} \right)^2 }
\label{eq:mean_TNE}
\end{equation}

Figure 3 (1)-(4) show the mean values of the TNE components at the interface obtained using \textit{type-c} particles. The period of $t=0$ to 2.5 is chosen for the study, and the interface particles track the interface shape better during the time. Comparing these four plots, we can see that most of the curves have peaks at $t=0.1$. At $t=0.1$, the interface has not yet evolved characteristic structures such as spikes and bubbles, and has not even started to evolve significantly. The fluid system keeps accumulating driving forces to drive the system to further destabilization. Figure 3 (1) shows the variation of the components of $\overline{\boldsymbol{\Delta}}^{*}_{2}$ with time, corresponding to the variation of the viscous stress tensor at the interface during the evolution. At $t=0.1$, $\overline{\boldsymbol{\Delta}}^{*}_{2,xx}$ and $\overline{\boldsymbol{\Delta}}^{*}_{2,yy}$ take their respective peaks, and then both gradually decrease. The curves of both are almost exactly antisymmetrically distributed along $y=0$. On the other hand, $\overline{\boldsymbol{\Delta}}^{*}_{2,xy}$, oscillates slowly with little change. The $\overline{\boldsymbol{\Delta}}^{*}_{3,yyy}$ and $\overline{\boldsymbol{\Delta}}^{*}_{3,xxy}$ also deviate from the equilibrium state from opposite places, although both are not very symmetric about the 0-axis.
In fact, $\overline{\boldsymbol{\Delta}}^{*}_{3,yyy}$+ $\overline{\boldsymbol{\Delta}}^{*}_{3,xxy}$ = $\overline{\boldsymbol{\Delta}}^{* }_{3,1,y}$. In the comparison with $\overline{\boldsymbol{\Delta}}^{*}_{3,1,x}$, $\overline{\boldsymbol{\Delta}}^{*}_{3,1,y}$ is always above the $\overline{\boldsymbol{\Delta}}^{* }_{31,x}$. To some extent it can be assumed that the vertical heat flow is always dominant compared to the horizontal heat flow. Only this dominance becomes weaker as time grows. The $\overline{\boldsymbol{\Delta}}^{*}_{4,2}$ represents a higher-order physical process. At $t=1.3$ $\overline{\boldsymbol{\Delta}}^{*}_{4,2,yy}$ surpasses $\overline{\boldsymbol{\Delta}}^{*}_{4,2,xx}$.  Different TNE components have different evolutionary patterns, and different TNE components have different magnitudes. As can be seen from Figure 3, $\overline{\boldsymbol{\Delta}}^{*}_{3}>\overline{\boldsymbol{\Delta}}^{*}_{3,1}>\overline{\boldsymbol{\Delta}}^{*}_{4,2}>\overline{\boldsymbol{\Delta}}^{*}_{2}$. The non-equilibrium phenomena are very rich and there is still much to be discovered. In general, a particular non-equilibrium state requires a comprehensive consideration of the various non-equilibrium components, which together paint a complete physical picture of the fluid system. The physical correspondence and interpretation of many of these components need to be further investigated.
\vspace*{4mm}
\begin{figure}
	\centerline{\includegraphics[width=16cm]{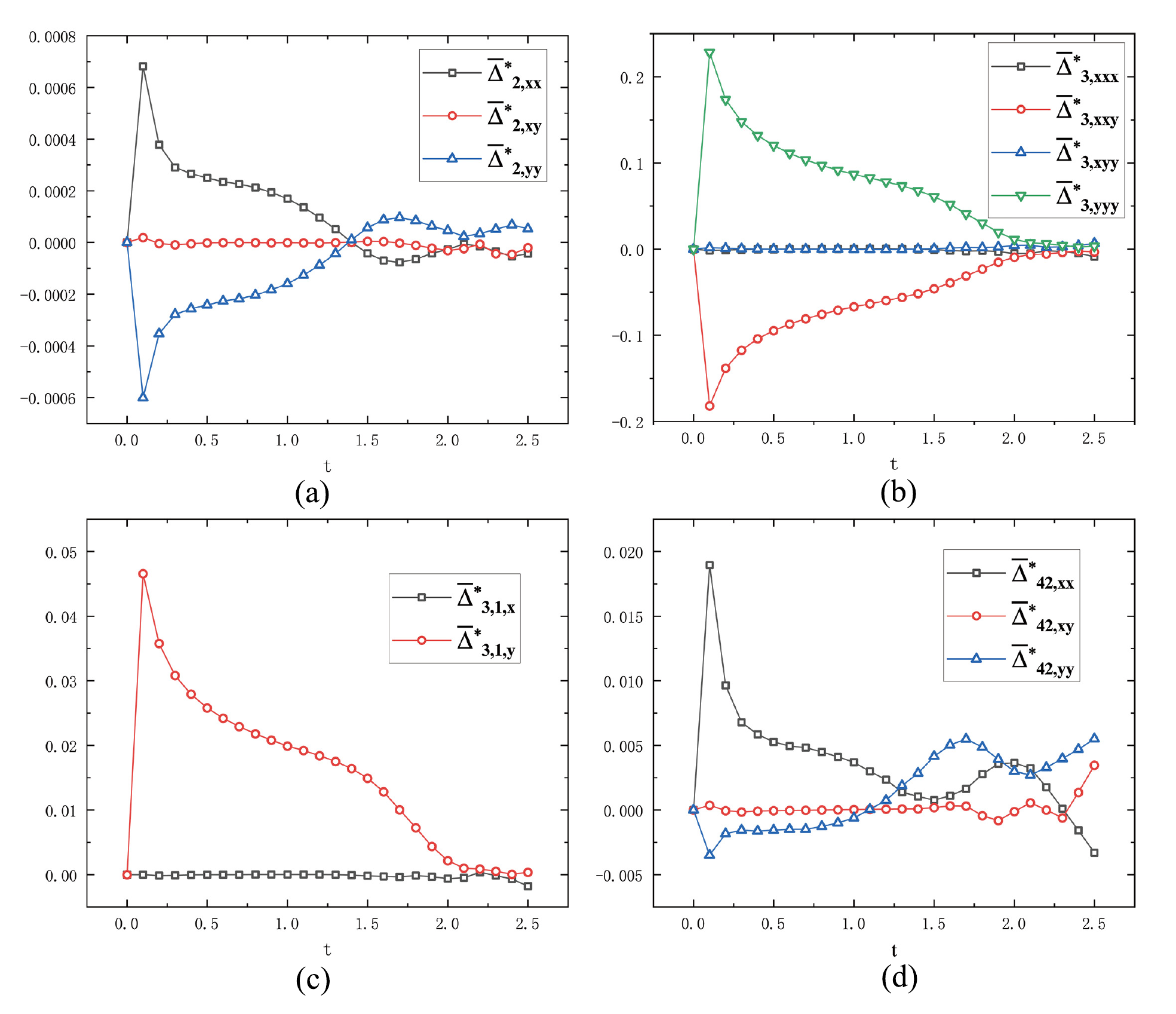}}
	\begin{center}
		\parbox{15.5cm}{\small{\bf Fig.3.}  Average value of TNE component of interface tracers. }
	\end{center}
\end{figure}
\vspace*{4mm}
\begin{figure}
	\centerline{\includegraphics[width=12cm]{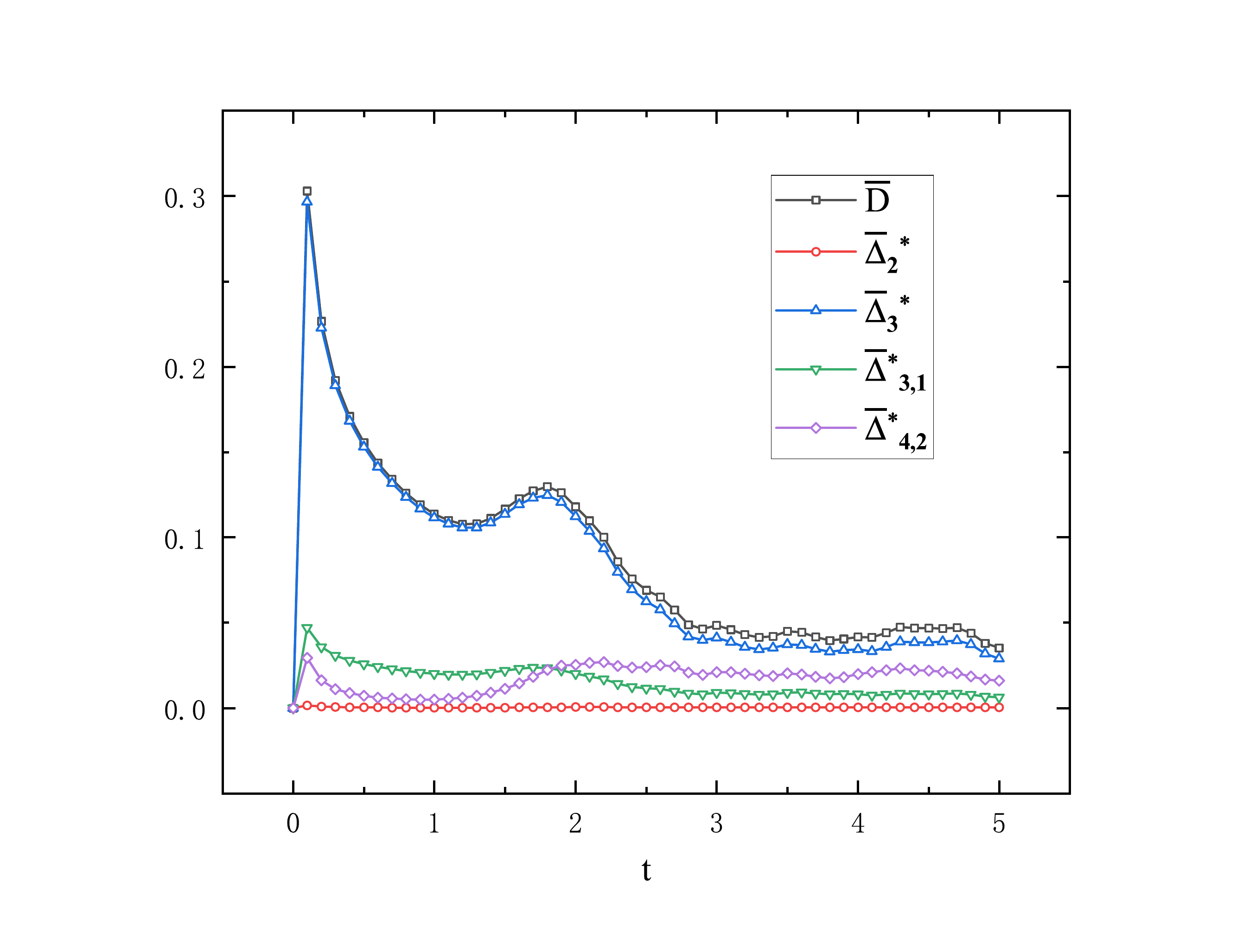}}
	\begin{center}
		\parbox{15.5cm}{\small{\bf Fig.4.}  Variation of mean TNE strength and its component with time. }
	\end{center}
\end{figure}

Figure 4 shows the variation of the mean TNE strength $\overline{D}$ at the interface with time, and the trend has similarity to the above component. Since the initial state given by the numerical simulation is not a steady state, the system is not in the state with the lowest degree of non-equilibrium at the initial moment. At the beginning of the simulation, TNE causes an initial destabilization of the system and causes the system to evolve spontaneously toward the equilibrium state. As the system dissipates and heat conduction occurs, the interface gradually becomes smooth and the transition layer becomes wider. The temperature and density gradient at the interface also decrease slowly due to diffusion. Therefore, $\overline{D}$ has a global maximum value at $t=0.1$ (this is the first time step of the statistics, when $\overline{D}$ is approximately equal to the initial system value) and decreases with time later. At $t=1.2$, $\overline{D}$ reaches a local minimum value and then increases, which corresponds to the start of significant growth of the perturbation interface under the effect of gravity. The width of the mixing zone is also growing rapidly, with the heavy fluid developing downward into spikes and the light fluid developing downward into bubbles. At $t=1.8$, $\overline{D}$ reaches a local maximum value during the growth and then slowly decreases, which means that the decrease in the interfacial macroscopic quantity gradient due to the growth of the interface length starts to dominate. As the interfacial shear triggers the occurrence of Kelvin–Helmholtz instability (KHI), vortex structures are generated on both sides of the spike and the bubble. The flow field starts to become more complex, and the mixing of light and heavy fluids gradually becomes more adequate. The interfacial density gradient and temperature gradient will naturally decrease significantly. The macroscopic quantity gradient can also characterize the TNE in one way. The rapid elongation of the interface makes the TNE increase, while the decrease of the macroscopic quantity gradient makes the TNE decrease, and the two compete with each other and work together to determine the growth trend of the mean non-equilibrium strength of the system at the interface.

\subsection{Viscous effects on mixing}\label{subsec:visco}
In the single relaxation time DBM with the ideal gas equation , the relaxation time $\tau$ together with the pressure $P$ determines the viscosity.
\begin{equation}
\mu=\tau \rho RT=\tau P
\label{eq:mu}
\end{equation}

The initial pressure $(p_{0}=1)$at the top of the upper part of the fluid  is used as the characteristic pressure, and the initial viscosity of the fluid can be changed by adjusting the relaxation time $\tau$. In addition, as an intrinsic physical parameter of the fluid, $\tau$ is related to the Knudsen number that characterizes the degree of non-equilibrium of the fluid system.

While describing complex flows, tracers also provide an additional way to quantify the degree of mixing of light and heavy fluids. Zhang, et al.$^{[38]}$ defined the local mixedness with the help of the location information of two types of tracers,
\begin{equation}
m_p \left (x,y \right )=4 \frac {n_a \left (x,y \right )\cdot n_b\left (x,y \right )}{\left [ n_a \left (x,y \right )+ n_b\left (x,y \right ) \right ]^2}
\label{eq:local_mixedness}
\end{equation}
where $n_a$ and $n_b$ are the local densities of \textit{type-a} particles and \textit{type-b} particles, respectively. If there are neither \textit{type-a} nor \textit{type-b} particles in a statistical cell, then the local mixing degree $m_p$ in this cell is the minimum value of 0. If the number of both particles in this statistical cell is equal, i.e., $n_a = n_b$, then $m_p$ is the maximum value of 1.0, which is considered as the fullest mixing at this time. The change of $m_p$ from 0 to 1 represents the gradual deepening of fluid mixing, from no mixing to complete mixing. The selection of the statistical cell is a key step. Here we choose statistical cells with twice the length of the grid size ($ds_x=n\cdot dx,ds_y=n\cdot dx,n=2$). Based on this, Zhang, et al. additionally defined the averaged mixedness (AM):
\begin{equation}
AM=\frac{1}{A}\int_{A} m_p \left (x,y \right ) d s_xd s_y
\label{eq:avm}
\end{equation}

The fluid mixing caused by multi-mode RTI at different viscosities is simulated, and Figure 5 (a) shows the variation of AM with time. From Figure 5 (a), it can be found that the AM curves of different viscosities almost coincide during early stage of mixing development, which indicates that the effect of viscosity is relatively weak at the early stage. With the growth of time, the AM of the low viscosity system grows faster, and its curve obviously lies above the AM curve of the high viscosity system. And the trend of AM decreasing with increasing initial viscosity coefficient is obvious, which indicates that the high viscosity effect inhibits mixing at the later stage. To explore the effect of viscosity on AM at the early stage of RTI development, we zoom in locally on the earlier part of the curve and obtain the information that the AM curves of the systems with different viscosities intersect in Figure 5 (b). This indicates that just like in single-mode RTI, the viscosity has a two-stage effect on fluid mixing induced by multi-mode RTI. There exists a characteristic time $t_0$. At the initial stage $(t<t_0)$, the AM of the high-viscosity fluid lies above that of the low-viscosity fluid, and viscosity can slightly enhance mixing; however, at the later stage $(t>t_0)$, the AM of the low-viscosity fluid is significantly larger, and high viscosity inhibits mixing.

Figure 6 shows the density contours and the tracer distribution patterns of the systems with different viscosities $(\mu_1=3.0\times10^{-5},\mu_2=7.0\times10^{-5},\mu_3=1.5\times10^{-4})$ at three different moments $(t=1.3,1.5,3.5)$. The three different moments correspond to the initial phase, the transitional phase and the late mixing phase, which can provide more detailed information to explain the viscous effects.

At $t=1.3$, as shown in Figure 6 (a1)(b1), the low-viscosity system presents more abundant small perturbations at the interface relative to the high-viscosity system. As the viscosity increases, these initial small perturbations merge into larger perturbations and the number of modes decreases. This is in general agreement with the findings of Liang, et al.$^{[23]}$ who studied the Reynolds number effect: a decrease in the Reynolds number of a fluid system promotes the merging of small perturbations. Thus, at the early stages, high viscosity promotes mixing mainly by dissipating small perturbations and promoting large structure evolution.

The $t=1.5$ is the turning point. Although the AM values of different systems are consistent at this time, the interface morphology shows obvious differences. At $t=3.5$, as shown in Figure 6 (a3)(b3), the fluid mixing is more adequate in the low-viscosity system and more small structures are generated in the systems, while the large structures remain relatively intact in the high-viscosity system. It can explain why the AM values of the low-viscosity system jump above those of the high-viscosity system from one perspective . On the other hand, as revealed by previous studies, KHI can effectively promote fluid mixing. The modes are more abundant in the low-viscosity system, and the development of KHI leads to modal coupling and increasingly strong interactions between the modes, which effectively contributes to the rapid growth of AM. In contrast, the number of modes in the high-viscosity system is reduced due to the merging of small perturbations, and the interactions between modes are weaker than those in the low-viscosity system. The AM curves of some systems in the later period seem to be not very different can also be explained from this perspective: if the number of modes and the structure are similar after the small perturbations merge, the first reason mainly works, and therefore the AM values show less obvious differences than the others.
\vspace*{4mm}
\begin{figure}
	\centerline{\includegraphics[width=16cm]{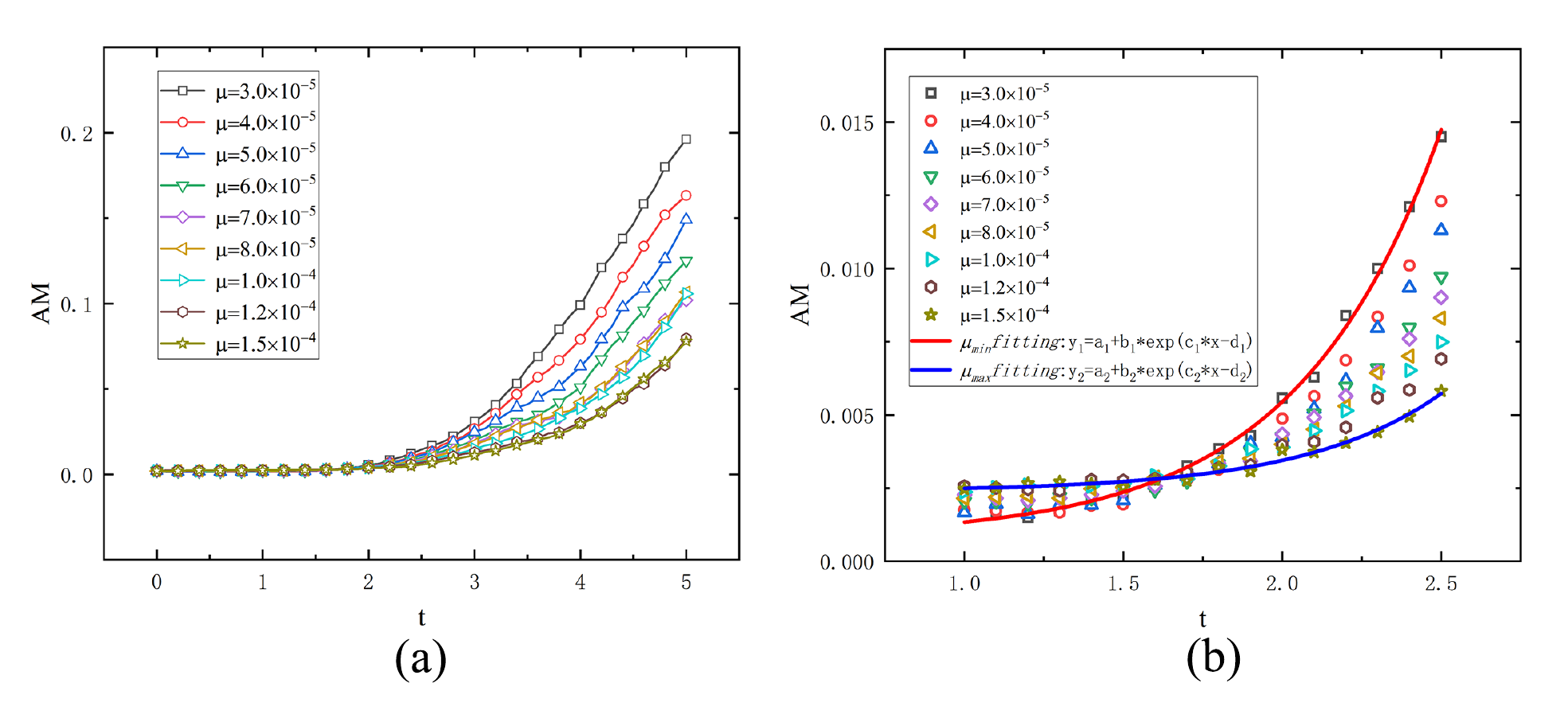}}
\begin{center}
	\parbox{15.5cm}{\small{\bf Fig.5.}  AM evolution of different viscous fluid systems with time; (a) Overall AM evolution diagram line; (b) Partial enlargement of AM curve with time $t$ from 1.0 to 2.5. }
\end{center}	
\end{figure}
\vspace*{4mm}
\begin{figure}
	\centerline{\includegraphics[width=16cm]{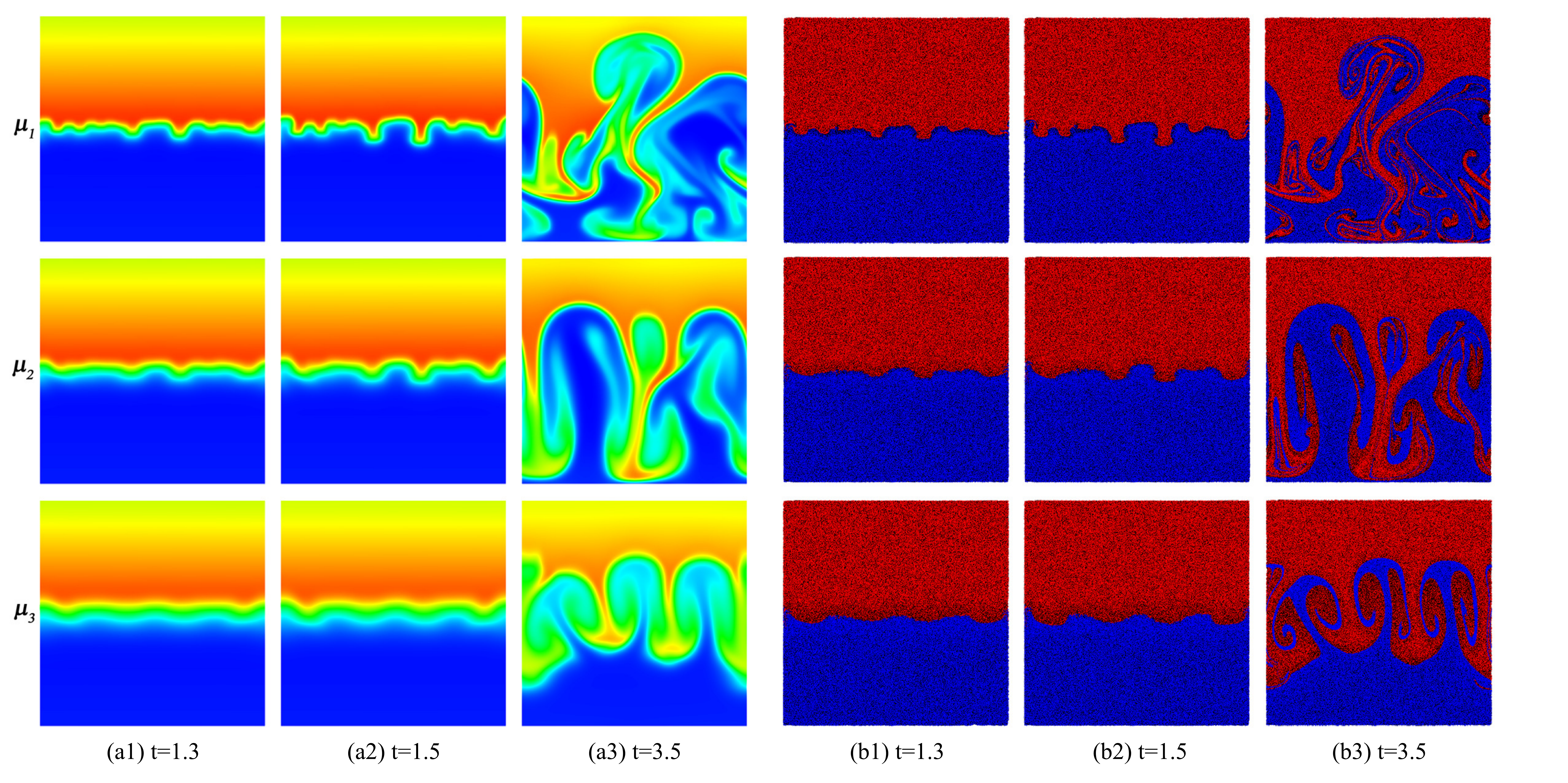}}
\begin{center}
	\parbox{15.5cm}{\small{\bf Fig.6.}  Density distribution and tracer particle images of fluid systems with different viscosities ($\mu_1=3.0\times10^{-5}, \mu_2=7.0\times10^{-5}, \mu_3=1.5\times10^{-4}$); (a) $t=1.3$; (b) $t=1.5$; (c) $t=3.5$. }
\end{center}	
\end{figure}
\vspace*{4mm}
\begin{figure}
	\centerline{\includegraphics[width=16cm]{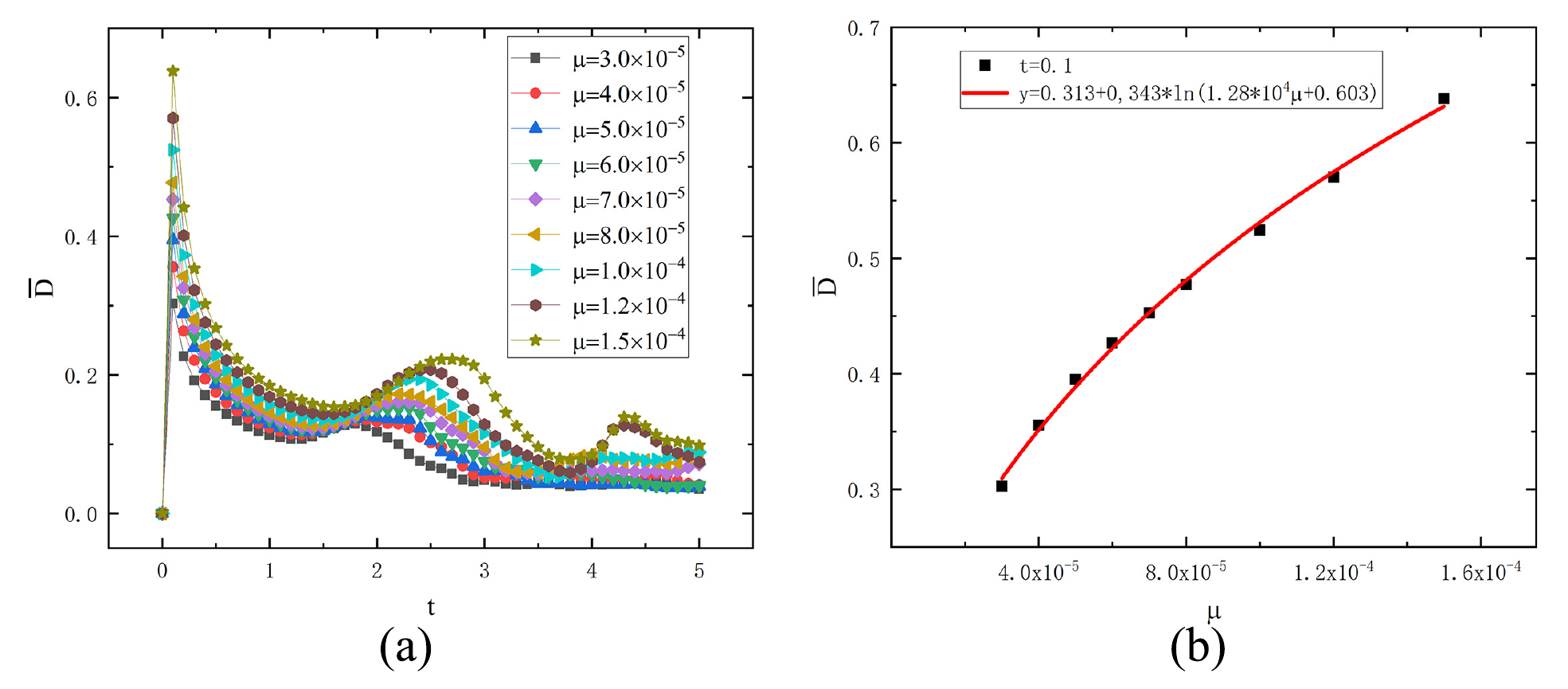}}
\begin{center}
	\parbox{15.5cm}{\small{\bf Fig.7.}  (a) Comparison of the mean TNE strength $\overline{D}$ evolution at the interface under different viscosities; (b) Effect of viscosity on $\overline{D}$ at $t=0.1$. }
\end{center}	
\end{figure}

Figure 7 (a) shows the evolution of the mean TNE strength $\overline{D}$ at the interface for different viscosities. Combined with the analysis of $\overline{D}$ under $\mu=3.0\times10^{-5}$, we can observe some interesting results. (1) The high-viscosity system has a larger value of $\overline{D}$ at $t=0.1$, i.e., the stronger the viscosity effect is, the higher the initial non-equilibrium degree of the interface. (2) The high-viscosity system takes longer to reach the local minimum value of $\overline{D}$, and the value is also larger at this time, which indicates that the higher the viscosity, the later the interface elongates rapidly and the interface possesses a higher degree of non-equilibrium at this time. (3) The high-viscosity system takes longer to reach the local maximum value of $\overline{D}$, and the value is also larger at this time, which indicates that the higher the viscosity, the later the interface macroscopic quantity gradient dominates. Overall, Figure 7 (a) can explain the evolution of fluid system with different viscosities from another perspective, where the higher the viscosity, the slower the rate of interfacial evolution. All physical processes are shifted back, although at similar physical states, highly-viscous fluid systems consistently have larger $\overline{D}$. Figure 7 (b) shows the relationship between the viscosity and the mean TNE strength of the interface at $t=0.1$, and $\overline{D}$ grows with $\mu$ in a logarithmic law.
\subsection{Effect of acceleration on mixing}\label{subsec:acceleration}
Figure 8 (a) shows the evolution of AM with time for different accelerations. It can be seen that the AM curves of the systems with different accelerations almost overlap during the early stage, indicating that the effect of acceleration on AM is smaller at the early stage. In the later period, the AM curves are separated, and the fluid system with higher acceleration is significantly more mixed. Figure 8 (b) shows the relationship between mixedness and acceleration for three characteristic moments ($t=3.5,t=4.0,t=4.5$) in the late mixing period. Fitting the data points using quadratic polynomials reveals that the AM grows faster with increasing acceleration. The role of acceleration in promoting mixing of light and heavy fluids is quite obvious, which seems to be well understood: with the increase of acceleration, the difference between gravity and buoyancy of heavy fluid also increases, so the spikes can insert into the bubbles more easily. But the tracer patterns tell us that there are other factors affecting this.
\vspace*{4mm}
\begin{figure}
	\centerline{\includegraphics[width=12cm]{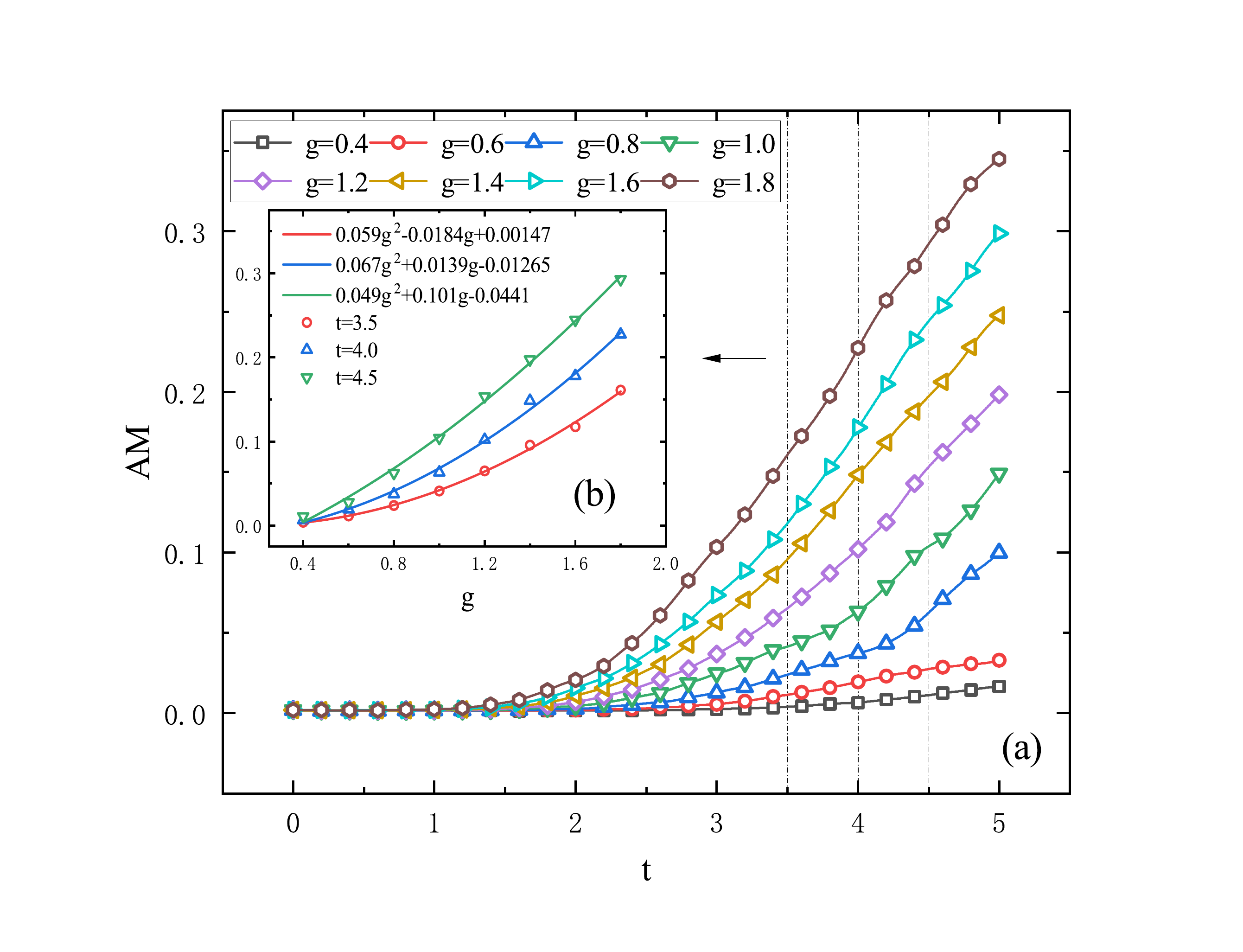}}
\begin{center}
	\parbox{15.5cm}{\small{\bf Fig.8.}  (a) Time evolution of AM in fluid systems with different acceleration; (b) the relationship between mixedness and acceleration at specific times. }
\end{center}	
\end{figure}
\vspace*{4mm}
\begin{figure}
	\centerline{\includegraphics[width=10cm]{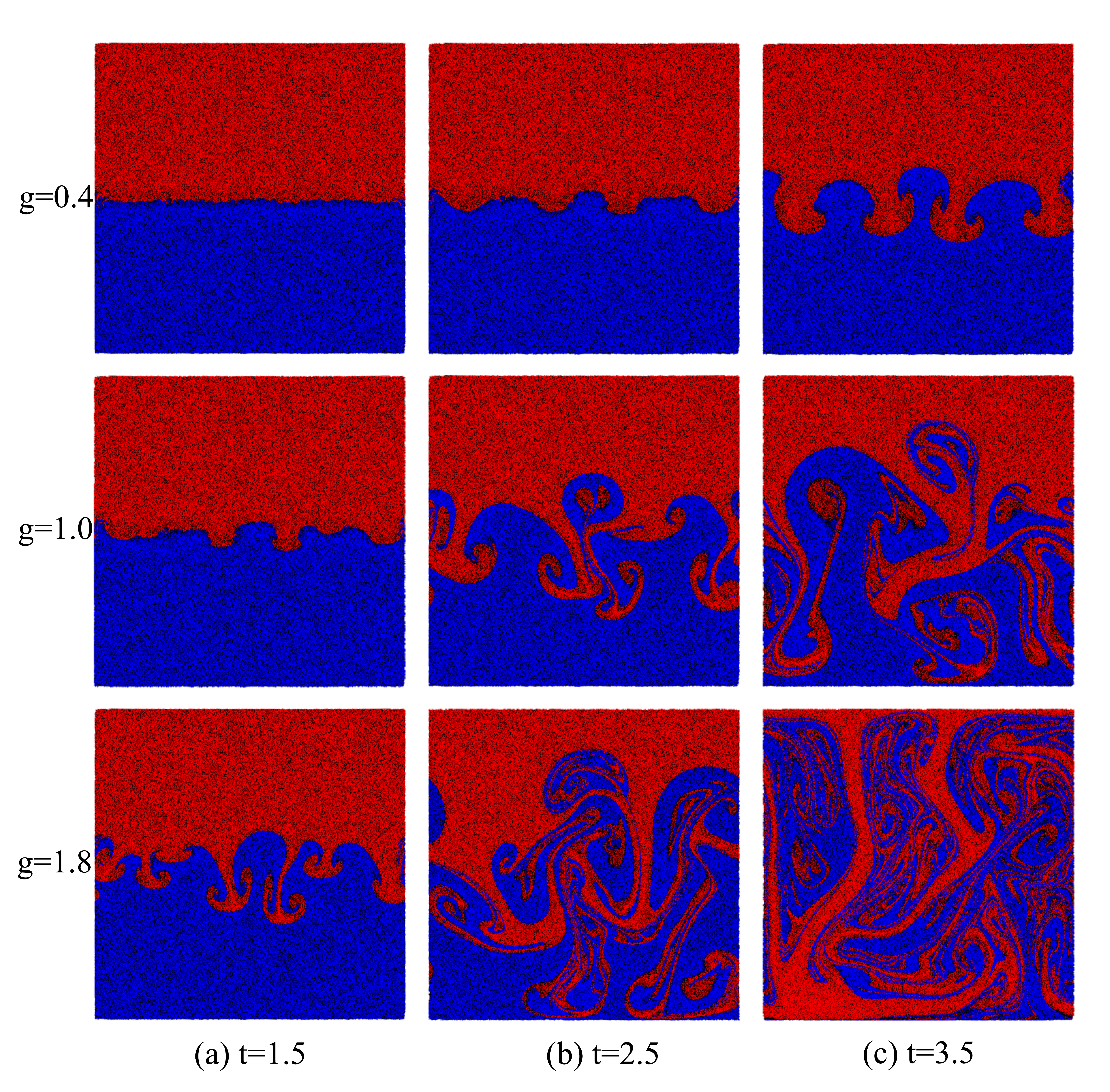}}
\begin{center}
	\parbox{15.5cm}{\small{\bf Fig.9.}  Tracer particles in fluid systems with different acceleration $(g_1=0.4, g_2=1.0, g_3=1.8)$; (a) $t=1.5$; (b) $t=2.5$; (c) $t=3.5$. }
\end{center}	
\end{figure}
\vspace*{4mm}
\begin{figure}
	\centerline{\includegraphics[width=16cm]{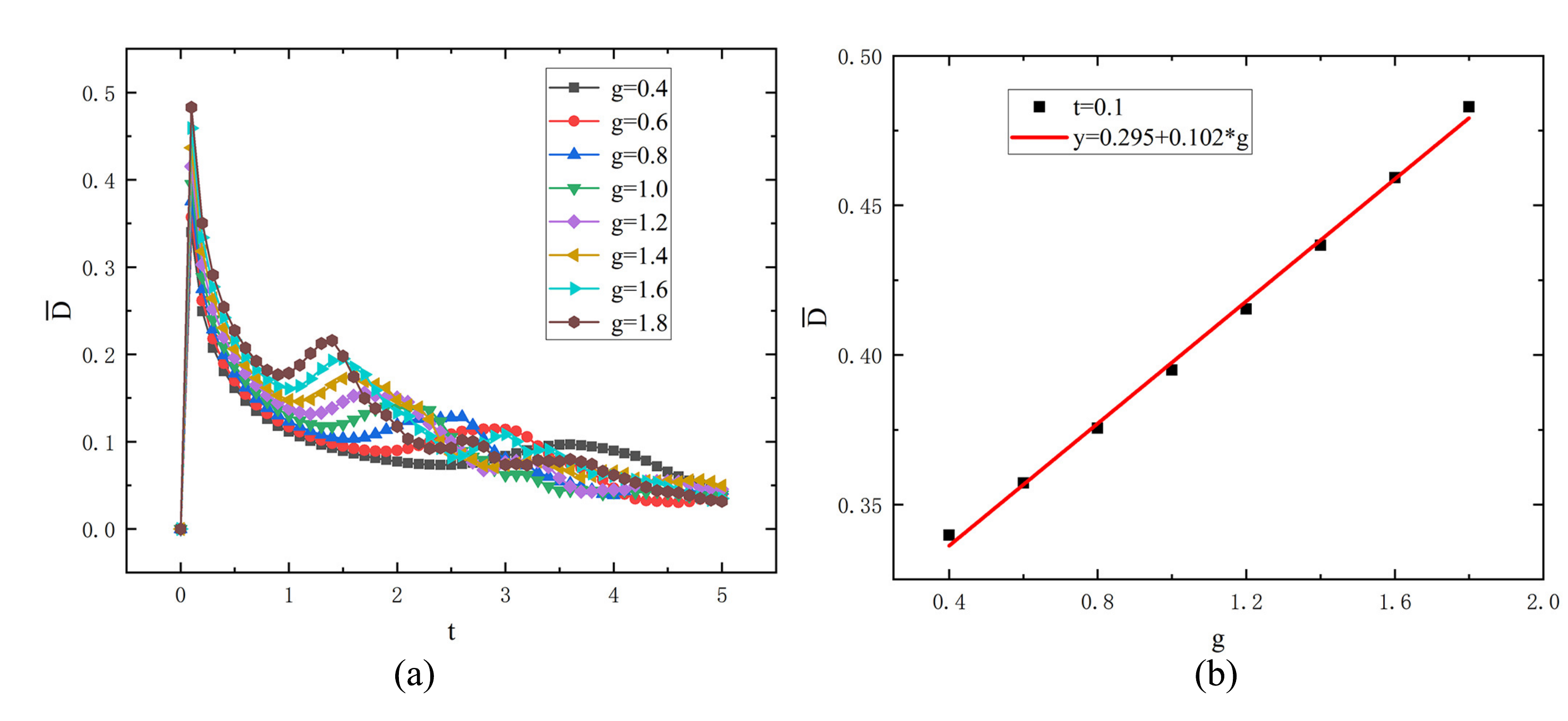}}
\begin{center}
	\parbox{15.5cm}{\small{\bf Fig.10.}  (a) Comparison of the mean TNE strength $\overline{D}$ evolution at the interface under different acceleration; (b) Effect of acceleration on $\overline{D}$ at $t=0.1$. }
\end{center}	
\end{figure}

As shown in Figure 9, the fluid system with higher acceleration has more abundant modes than the fluid system with lower acceleration. More small perturbations develop into spikes and bubbles in the system with large acceleration. These rich modal interactions and couplings strongly enhance fluid mixing. In the system with small acceleration, small pertubations are combined into a small number of large pertubations, and the mixing of light and heavy fluids is slower, which is another reason why acceleration promotes fluid mixing.

Figure 10 (a) shows the variation of mean TNE strength $\overline{D}$ at the interface for different accelerations. Combined with the previous analysis, some results are obvious. (1) The stronger the acceleration effect is, the higher the initial non-equilibrium degree of the interface and the stronger the tendency to evolve to the equilibrium state. As shown in Figure 10 (b), $\overline{D}$ grows linearly with the growth of $g$. (2) The system with higher acceleration takes less time to reach the local minimum value of $\overline{D}$, and the interface length starts to grow significantly earlier. The interface possesses a higher degree of non-equilibrium during the time; (3) The system with higher acceleration takes less time to reach the local maximum value, and the value is also larger, which indicates that the stronger the acceleration effect, the earlier the interface macroscopic quantity gradient dominates. Overall, the higher the acceleration, the faster the interface evolves. Physical processes move forward and systems with higher acceleration always have larger $\overline{D}$ in similar physical states.
\subsection{Effect of compressibility on mixing}\label{subsec:compressibility}
\vspace*{4mm}
\begin{figure}
	\centerline{\includegraphics[width=12cm]{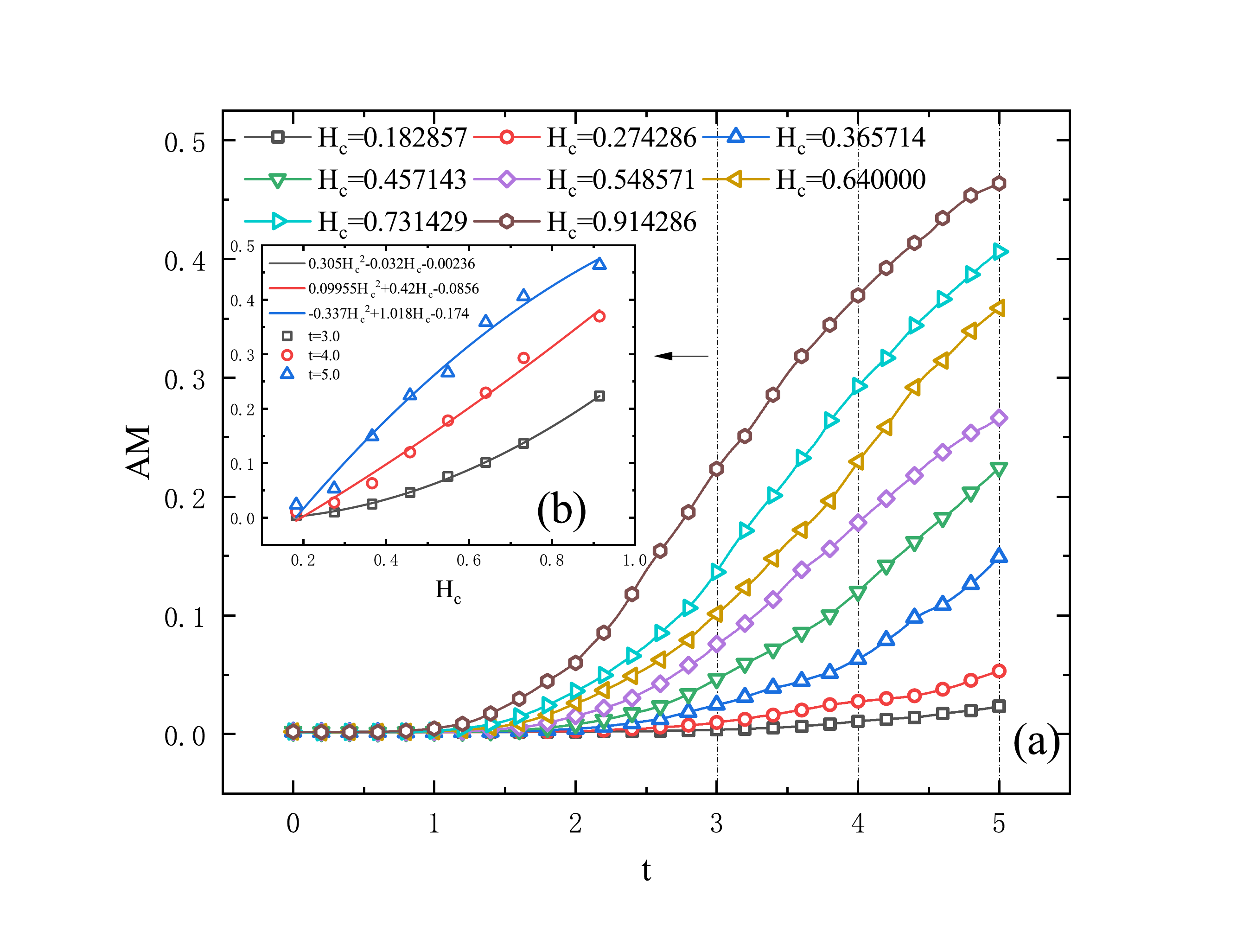}}
\begin{center}
	\parbox{15.5cm}{\small{\bf Fig.11.}  (a) Time evolution of AM in fluid systems with different compressibility; (b) the relationship between mixedness and compressibility at specific times. }
\end{center}	
\end{figure}
\vspace*{4mm}
\begin{figure}
	\centerline{\includegraphics[width=10cm]{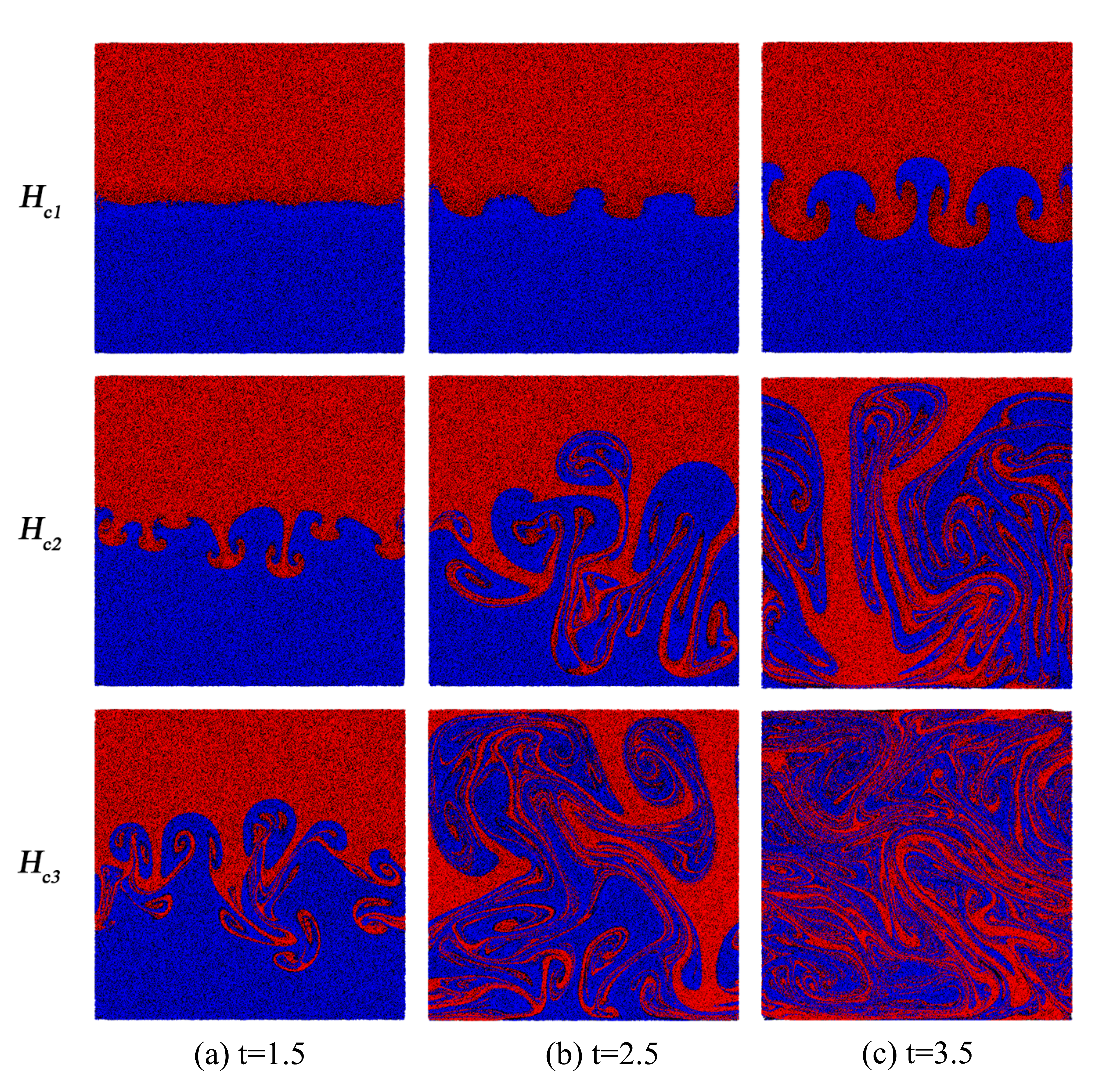}}
\begin{center}
	\parbox{15.5cm}{\small{\bf Fig.12.}  Tracer particles in fluid systems with different compressibility ($H_{c1}=0.182857, H_{c2}=0.548571, H_{c3}=0.914286$); (a) $t=1.5$; (b) $t=2.5$; (c) $t=3.5$. }
\end{center}	
\end{figure}
\vspace*{4mm}
\begin{figure}
	\centerline{\includegraphics[width=16cm]{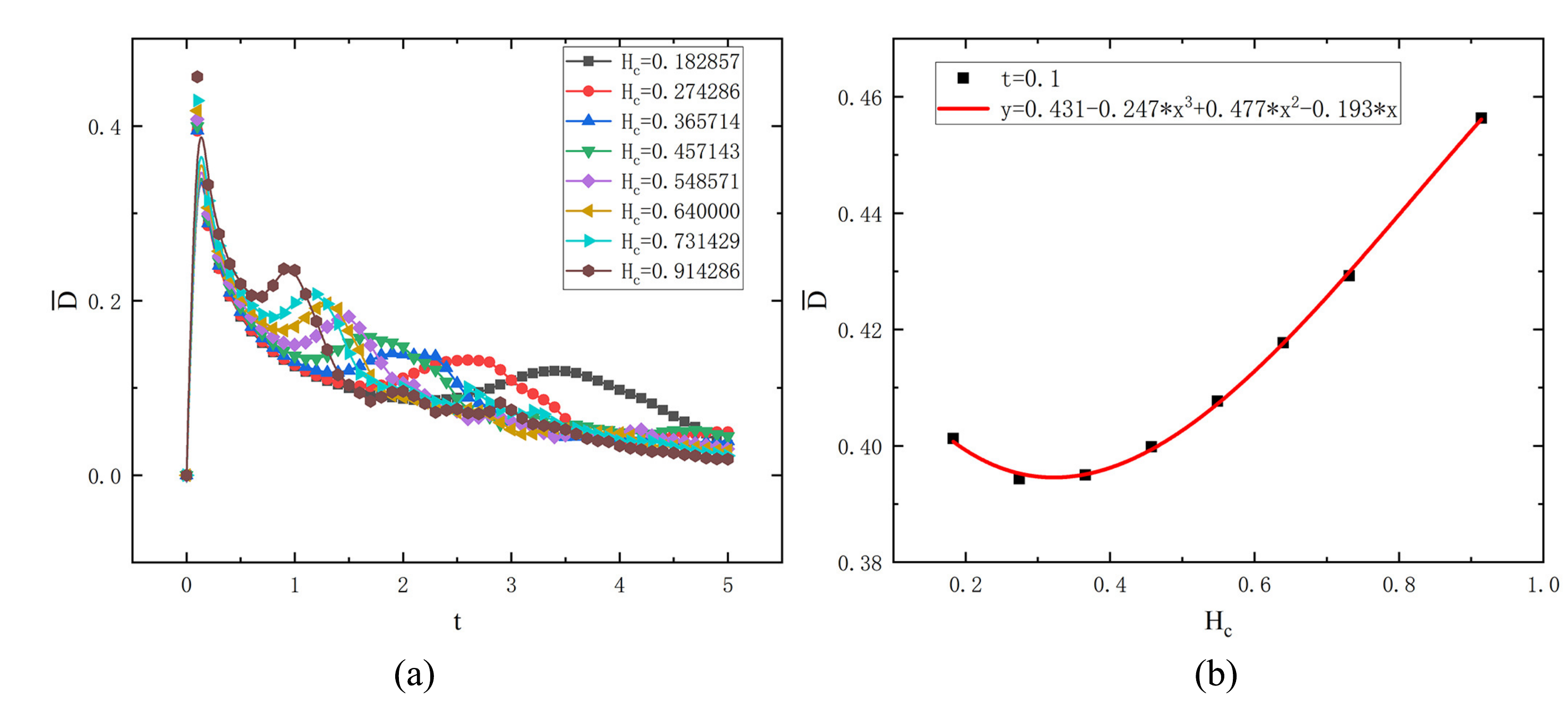}}
\begin{center}
	\parbox{15.5cm}{\small{\bf Fig.13.}  (a) Comparison of the mean TNE strength $\overline{D}$ evolution at the interface under different compressibility; (b) Effect of compressibility on $\overline{D}$ at $t=0.1$. }
\end{center}	
\end{figure}

Lai, et al. and Zhang, et al.$^{[29,38]}$ defined two dimensionless quantities $H_c$ and $H_\tau$ in their study of compressibility effects. Here we replace the wave number $1/k$ with the simulated region width $L_x$ as the characteristic length, and we can obtain the formulae for these two dimensionless numbers as follows.
\begin{equation}
H_c=\left(\frac{\sqrt{g\cdot L_x}}{c_s} \right)^2,\\
H_\tau=\frac{\tau}{\left (g/L_x \right)^{-1/2}}\\
\label{eq_H_c}
\end{equation}
where $H_c$ characterizes the compressibility and $H_\tau$ characterizes the relative non-equilibrium degree of the system. When exploring the effect of compressibility, it is common to adjust $\tau$ and $g$ in order not to affect the characteristic scales of different compressible fluid systems, and to change the compressibility coefficient $H_c$ without changing $H_\tau$.

Figure 11 (a) shows the variation of AM with time for different compressibility factors. Somewhat similar to the effect of acceleration, the effect of compressibility during the early stage is relatively small and the AM curves stack up. The AM value of the high-compressibility system grows faster at the later stage, indicating that the strong compressibility can significantly enhance the mixing at the later stage. Figure 11 (b) shows the relationship between the mixedness and the compressibility for three characteristic moments ($t=3.0,t=4.0,t=5.0$) at the late mixing stage. It can be found that although the AM increases with the compressibility coefficient at different moments, the growth trend is different. In particular, at $t=5.0$, the fitted curve tends to slow down and the AM tends to a saturation value. Combined with Figure 12 we can know that this is because at this time the highly compressible system is more fully mixed, the mixing degree gradually approaching saturation. The specific reason why high compressibility promotes mixing we can get some explanation from Figure 12.

Figure 12 shows tracer distribution patterns of the fluid system for three different moments $(t=1.5,2.5,3.5)$ with different compressibility coefficients $(H_{c1}=0.182857,H_{c2}=0.548571,H_{c3}=0.914286)$. From the figure, we can observe that the large fluid structure of the low-compressible system is relatively intact. As the compressibility factor increases, the interface deformation is more intense at the same moment, and the system produces more abundant small decomposition. In addition, like acceleration effects, highly compressible fluid systems have more complex modes, with small perturbations more likely to develop into spike-bubble structures and more intense interactions between modes. Moreover, highly compressible fluid systems have greater compression energy and a stronger tendency to release compression energy to reach an energy-minimizing equilibrium state.

Figure 13 (a) shows the evolution of the mean TNE strength $\overline{D}$ at the interface for different compressibility, which has some differences from the previous results. (1) The $\overline{D}$ decreases and then increases with the compressibility coefficient at the $t=0.1$. As shown in Figure 13 (b), it varies with a cubic polynomial law. (2) The larger the compressibility coefficient, the earlier the interface length starts to grow significantly and the interface possesses a higher degree of non-equilibrium at this time. (3) The larger the compressibility coefficient, the earlier the interface macroscopic quantity gradient dominates.  Overall, similar to the acceleration effect, compressibility effectively promotes the evolution speed of interface. The compressibility can be changed by adjusting $\tau$ and $g$. In order to make $H_c$ increase while keeping $H_\tau$ constant, it is necessary to increase $g$ and decrease $\tau$ at the same time. The two compete with each other and jointly determine the evolution of different compressibility systems. Because the highly compressible system has both larger acceleration and lower viscosity, the interface evolves faster than one factor alone. The initial non-equilibrium of the interface also decreases and then increases with increasing compressibility factor because these two factors compete with each other.
\subsection{Effect of Atwood number on mixing}\label{subsec:At}
\vspace*{4mm}
\begin{figure}
	\centerline{\includegraphics[width=10cm]{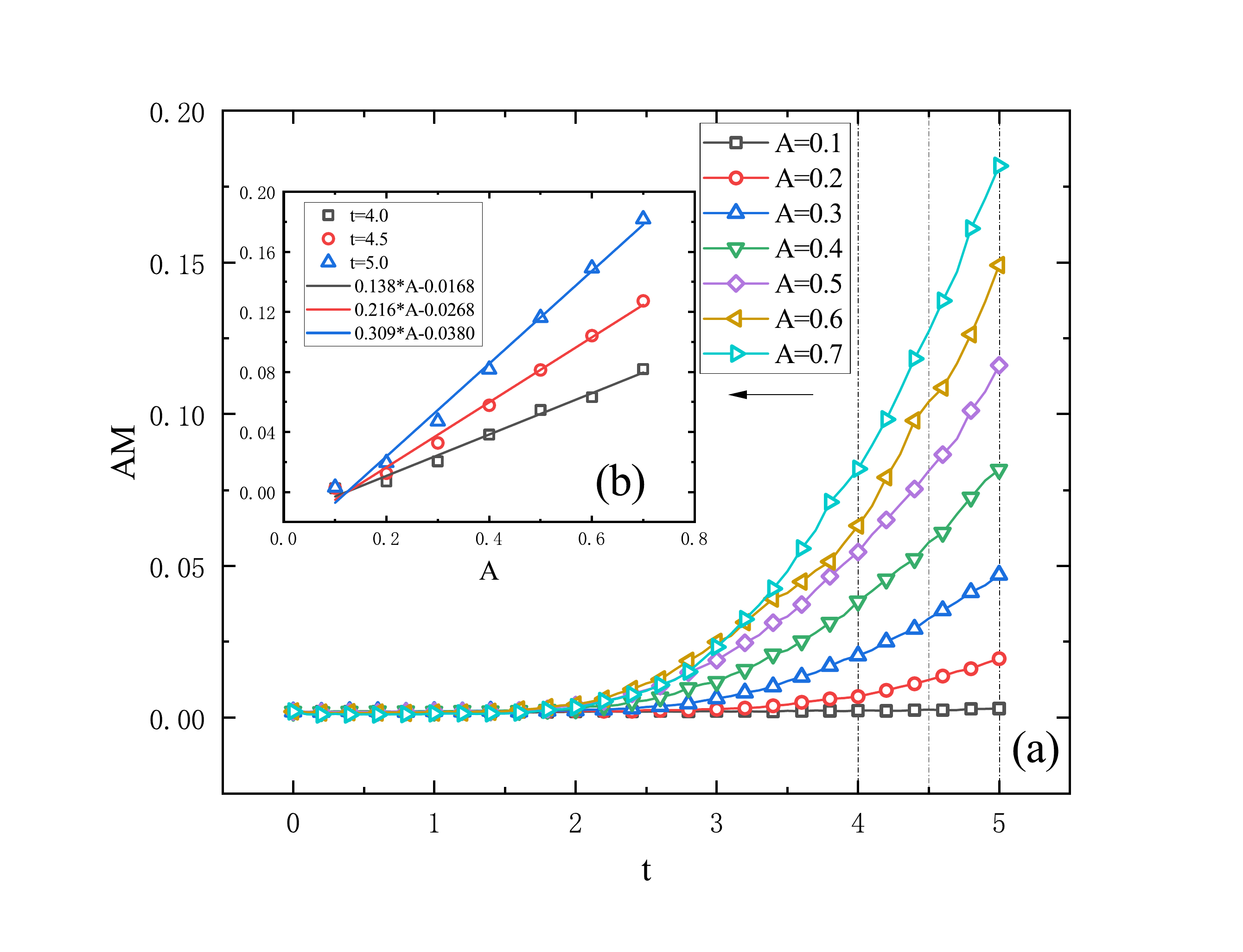}}
\begin{center}
	\parbox{15.5cm}{\small{\bf Fig.14.}  (a) Time evolution of AM in fluid systems with different Atwood number; (b) the relationship between mixedness and Atwood number at specific times. }
\end{center}	
\end{figure}
\vspace*{4mm}
\begin{figure}
	\centerline{\includegraphics[width=8cm]{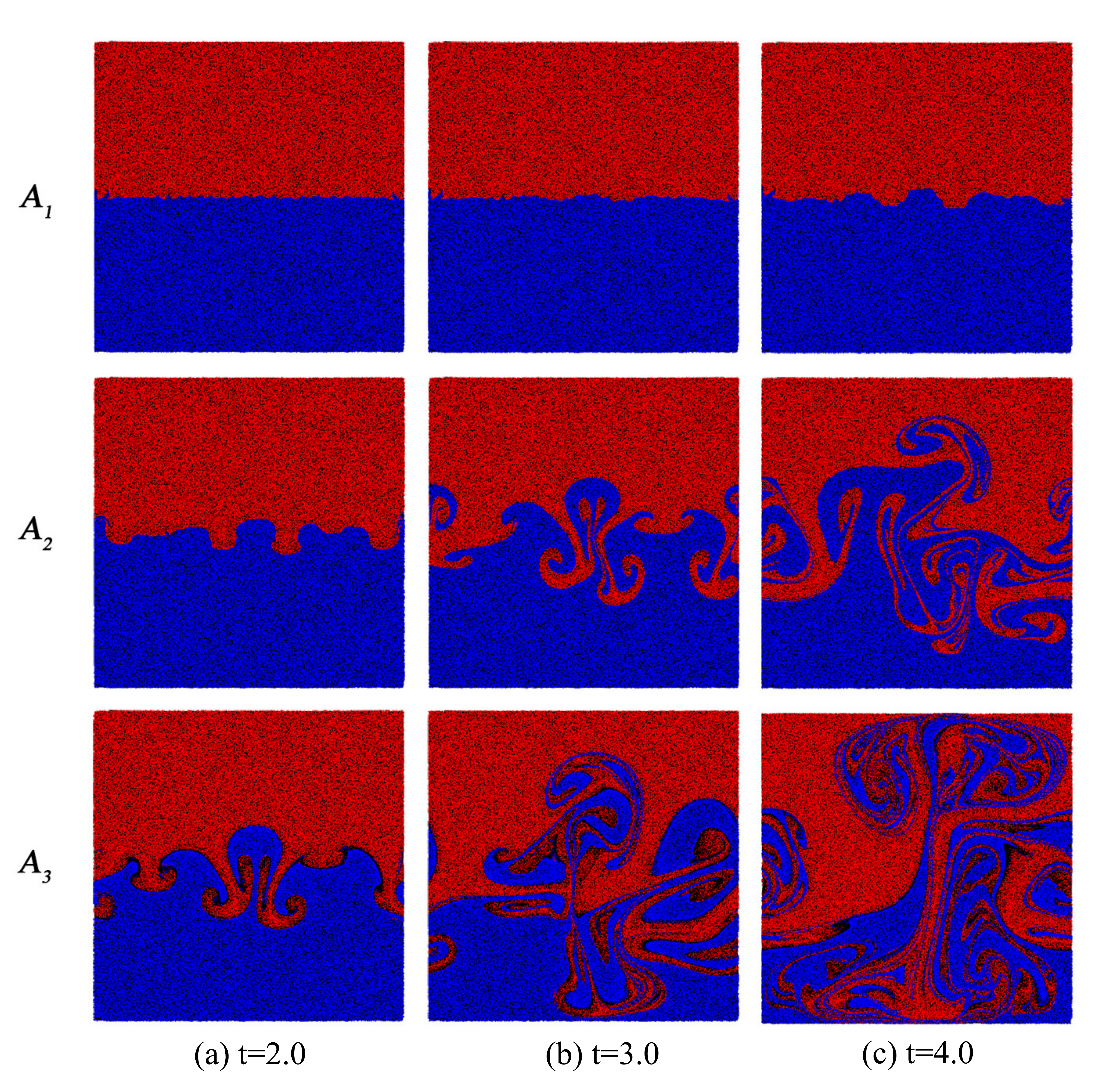}}
\begin{center}
	\parbox{15.5cm}{\small{\bf Fig.15.}  Tracer particles in fluid systems with different Atwood number ($A_1=0.1, A_2=0.4, A_3=0.7$); (a) $t=2.0$; (b) $t=3.0$; (c) $t=4.0$. }
\end{center}	
\end{figure}
\vspace*{4mm}
\begin{figure}
	\centerline{\includegraphics[width=16cm]{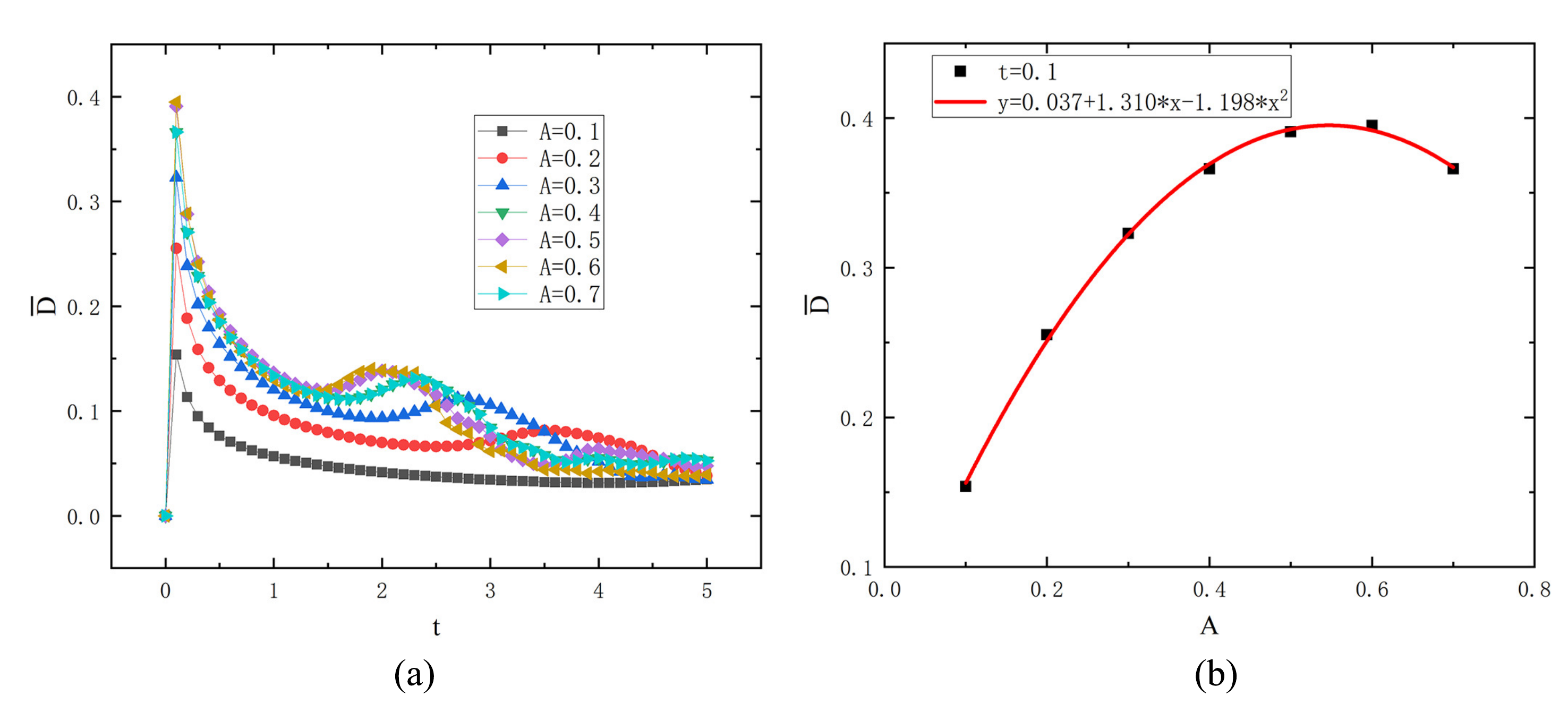}}
\begin{center}
	\parbox{15.5cm}{\small{\bf Fig.16.}  (a) Comparison of the mean TNE strength $\overline{D}$ evolution at the interface under different Atwood number; (b) Effect of Atwood number on $\overline{D}$ at $t=0.1$. }
\end{center}	
\end{figure}

Figure 14 (a) illustrates the evolution of AM for systems with different Atwood numbers. It is easy to see from the figure that the Atwood number has less effect at the early stages; at the later stages, a higher Atwood number can effectively promote mixing. Figure 14 (b) selects three characteristic moments ($t=4.0,t=4.5,t=5.0$) to study the relationship between the mixedness and the Atwood number, and finds that they are linearly correlated and the slope increases gradually with time. To further explain this phenomenon, Figure 15 gives tracer patterns of the systems with different Atwood numbers ($A_1=0.1, A_2=0.4, A_3=0.7$) at three different moments ($t=2.0, 3.0, 4.0$). It can be observed that the Atwood number does not promote perturbation merging, which is different from the other three factors. The higher Atwood number promotes mixing mainly by increasing the difference between gravity and buoyancy of heavy fluid and increasing the temperature and density gradients to promote thermal and mass diffusions because single fluid model is adopted.

Figure 16 (a) shows the evolution of the mean TNE strength $\overline{D}$ at the interface for different Atwood number. (1) The $\overline{D}$ at the initial moment ($t=0.1$) increases and then decreases with the Atwood number. As shown in Figure 16 (b),  it changes in a parabolic law. (2) The turning point when the interface length starts to grow significantly and the moment when the macroscopic quantity gradient at the interface dominates are also advanced and then delayed with the Atwood number. This seems counter-intuitive, but in fact the equilibrium state of the fluid system is related to the initial state setting. The non-equilibrium characterized by $\overline{D}$ is actually a deviation of the system relative to its respective equilibrium state, which is inherently different for systems with different Atwood number. Intuitively it seems that the larger the Atwood number, i.e., the larger the density difference, the larger the non-equilibrium of the system, and Figure 16 (b) illustrates that they are not necessarily positively correlated.

\section{Conclusions}\label{sec:conclusions}

In this paper, multi-mode perturbation RTI is investigated by coupling tracers with DBM. The physical functions of tracers are further extended. Besides the position space probed previously, we explore further the position-velocity phase space where the position and velocity information carried by the tracers constitutes a series of fascinating patterns and evolution schemes. The phase space provides a new intuitive geometric observation to the complex flow behaviors, which is helpful for the future in-depth study of complex fluid systems.

Various types of non-equilibrium behaviors at interfaces are investigated using interfacial particle tracking techniques. The interface length and the interface macroscopic quantity gradient compete with each other to jointly determine the mean TNE strength at interfaces. The effects of physical factors such as viscosity, acceleration, compressibility, and Atwood number on the mixings of matter and momentum, and the mean TNE strength at the interface are investigated. It is found that the mixedness increases gradually with increasing viscosity during the early stage, but decreases with increasing viscosity at a later stage. At the early stage, the high viscosity enhances mixing mainly by dissipating small perturbations and promoting large structure evolution, while it inhibits mixing by withholding the generation of small structures and weakening the interactions among different vortexes at the later stage. Larger acceleration promotes mixing more significantly. With the increase of gravity acceleration, the difference between gravity and buoyancy of heavy fluid also increases, consequently it becomes easier for the heavy fluid to insert into the light fluid. During this process, a larger gravity acceleration induces larger shear velocities at the interface which are the origins of Kelvin-Helmholtz vortex. Moreover, being different from the case of single-mode perturbation, a larger gravity acceleration favors the development of small wavelength perturbations. Compressibility promotes mixing mainly by favoring generating and developing small structures, as well as favoring generating richer perturbation modes and interactions between later modes. Higher Atwood number favors the RTI evolution by making the gravity of the heavy fluid more remarkable with respect to the buoyancy from the light fluid. Since single fluid model is adopted in this paper, the temperature and density gradients increase with higher Atwood number, corresponding to higher thermal and mass diffusions.

The stronger the viscous effect, the higher the initial interfacial non-equilibrium degree, the later the interface length grows significantly, and the later the macroscopic quantity gradient dominates. The greater the acceleration, the higher the initial interface non-equilibrium degree, the earlier the interface length grows significantly, and the earlier the macroscopic quantity gradient dominates. The stronger the compressibility, the earlier the time for rapid elongation of the perturbed interface, the earlier the macroscopic quantity gradient dominates, but the initial non-equilibrium degree of the interface decreases first and then increases. In contrast, as the Atwood number increases, the initial non-equilibrium degree of the interface first increases and then decreases; and both of the turning point for interface length dominating and the turning point for macroscopic quantity gradient dominating first become earlier and then are delayed.

\section*{Acknowledgments}
Thanks to Huilin Lai, Jie Chen, Dejia Zhang, Jiahui Song, and Yanbiao Gan for their kind help. This work was supported by the National Natural Science Foundation of China (under Grant No. 12172061) and the Opening Project of State Key Laboratory of Explosion Science and Technology (Beijing Institute of Technology) under Grant No. KFJJ21-16 M.

\end{document}